\documentclass[prb,twocolumn,superscriptaddress]{revtex4-1}

\usepackage[dvips]{graphicx}
\usepackage{epsfig}

\usepackage[margin=2cm,head=0.5cm]{geometry}
\usepackage[latin1]{inputenc}
\usepackage{bm}
\usepackage{amsmath}
\usepackage{amssymb}
\usepackage{latexsym}
\usepackage{amsfonts}
\usepackage{epsfig}
\usepackage[dvipsnames]{xcolor}
\usepackage[linktocpage, colorlinks=true ,linkcolor=blue, citecolor=blue]{hyperref}
\usepackage[all]{hypcap}

\newcommand{\expval}[1]{\left< #1 \right>}

\newcommand{\ket}[1]{\left|#1\right>}
\newcommand{\bra}[1]{\left<#1\right|}

\newcommand{\matel}[3]{\left<#1\left|#2\right|#3\right>}
\newcommand{\nn}{\nonumber\\}

\newcommand{\f}[1]{\mbox{\boldmath$#1$}}

\newcommand{\ord}{{\cal O}}

\newcommand{\traceB}[1]{{\rm Tr_B}\left\{ #1 \right\}}

\newcommand{\ptrace}[2]{{\rm Tr_{#1}}\left\{ #2 \right\}}
\newcommand{\abs}[1]{{\left| #1 \right|}}

\newcommand{\ii}{\mathrm{i}}  

\newcommand{\dn}{{\mathord{\downarrow}}}
\newcommand{\up}{{\mathord{\uparrow}}}
\newcommand{\new}[1]{{#1}}

\begin{document}
  
\title{Environment-induced~decay~dynamics~of~anti-ferromagnetic~order
in~\new{Mott-Hubbard~systems}}

\begin{abstract}
We study the dissipative Fermi-Hubbard model in the limit of weak tunneling and strong 
repulsive 
interactions, where each lattice site is tunnel-coupled to a Markovian fermionic bath. 
For cold baths at intermediate chemical potentials, the Mott insulator property 
remains stable and we find a fast relaxation of the particle number towards half filling. 
On longer time scales, we find that the anti-ferromagnetic order of the Mott-N\'eel ground 
state on bi-partite lattices decays, even at zero temperature. 
For zero and non-zero temperatures, we 
quantify the different relaxation time scales by means of waiting time distributions 
which can be derived from an effective (non-Hermitian) Hamiltonian and obtain fully analytic 
expressions for the Fermi-Hubbard model on a tetramer ring. 
\end{abstract}

\author{G.~Schaller}
\affiliation{Helmholtz-Zentrum Dresden-Rossendorf, Bautzner Landstra{\ss}e 400, 01328 Dresden, Germany}

\author{F.~Queisser}
\affiliation{Helmholtz-Zentrum Dresden-Rossendorf, Bautzner Landstra{\ss}e 400, 01328 Dresden, Germany}
\affiliation{Institut f\"ur Theoretische Physik, Technische Universit\"at Dresden, 01062 Dresden, Germany}

\author{N.~Szpak}
\affiliation{Fakult\"at f\"ur Physik and CENIDE, Universit\"at Duisburg-Essen, Lotharstra{\ss}e 1, 47057 Duisburg, Germany}

\author{J.~K\"onig}
\affiliation{Fakult\"at f\"ur Physik and CENIDE, Universit\"at Duisburg-Essen, Lotharstra{\ss}e 1, 47057 Duisburg, Germany}

\author{R.~Sch\"utzhold}
\affiliation{Helmholtz-Zentrum Dresden-Rossendorf, Bautzner Landstra{\ss}e 400, 01328 Dresden, Germany}
\affiliation{Institut f\"ur Theoretische Physik, Technische Universit\"at Dresden, 01062 Dresden, Germany}

\date{\today}

\maketitle

\section{Introduction}

An important question in non-equilibrium physics of quantum many-body systems is the relaxation 
towards equilibrium after the excitation by an external stimulus~\cite{Kollath2007,Eckstein2009,moeckel2008a,trotzky2012a}.
For weakly interacting quantum many-body systems, there has been considerable progress in this 
direction because such systems often admit an effective single-particle description, e.g. via linearization around a mean field~\cite{bruus2002,lacroix2014a}.
For strongly interacting quantum many-body systems however, our understanding 
-- although advanced in one-dimensional systems~\cite{Imada1998,Lieb2004,essler2005,Edegger2007,Prosen2014,nakagawa2021a,bertini2021a} 
-- in general is still far from complete.

As one of the most prominent examples for a strongly interacting quantum many-body system,
we consider the Fermi-Hubbard Hamiltionian~\cite{Hubbard1963,tasaki1998a,jaksch2005,essler2005}
\begin{align}
\label{Fermi-Hubbard}
\hat H
&=
-J\sum_{\langle\mu\nu\rangle,s}\hat c^\dagger_{\mu,s}\hat c_{\nu,s}
+U\sum_\mu \hat n_\mu^\uparrow\hat n_\mu^\downarrow 
+\epsilon\sum_{\mu,s}\hat n_\mu^s
\nn
&=
\hat H_J+\hat H_U+\hat H_\epsilon
\,,
\end{align}
where $\hat c^\dagger_{\mu,s}$ and $\hat c_{\nu,s}$ denote the fermionic creation and 
annihilation operators at the lattice sites $\mu$ and $\nu$ with spin 
$s\in\{\uparrow,\downarrow\}$, while $\hat n_\mu^s=\hat c^\dagger_{\mu,s}\hat c_{\mu,s}$
is the corresponding number operator. 
The hopping strength $J$ describes tunneling between neighboring lattice sites 
$\langle\mu\nu\rangle$ and is supposed to be much smaller than the on-site repulsion 
or interaction strength $U$. 
Finally, we included the single-particle on-site energy $\epsilon$. 

In the case of half filling, 
the ground state of the Hubbard model~\eqref{Fermi-Hubbard} 
in higher dimensions would be metallic for weak interactions $U\ll J$ but it becomes insulating for strong repulsion $U\gg J$~\cite{Schaefer2015}. 
%
This Mott-insulating state is separated by the Mott gap $\approx U$ from those excited states containing doublon-holon pairs and
has mostly one particle per lattice site, 
but e.g., also features a small double occupancy 
$\langle\hat n_\mu^\uparrow\hat n_\mu^\downarrow\rangle\sim J^2/U^2$ due to (virtual) hopping processes which lower the energy~\cite{cleveland1976a}. 
At half filling, these hopping processes are only allowed for opposite spins at 
neighboring sites and thus induce an effective anti-ferromagnetic interaction. 
As a result, the ground state displays anti-ferromagnetic order (Mott-N\'eel state)
on bi-partite lattices (in higher dimensions).

Variants of the Hubbard model are investigated for signs of superconductivity~\cite{klein2006a,Qin2020}.
This has sparked tremendous efforts, such that nowadays experimental realizations 
of the Fermi-Hubbard model~\eqref{Fermi-Hubbard} include 
ultra-cold fermionic atoms in optical lattices~\cite{hofstetter2002a,liu2004a,Esslinger2010,Hart2015}
as well as electrons in various lattice systems, for example  
ad-atoms on Si surfaces~\cite{Hamers1988,salfi2016a}, 
arrays of quantum dots~\cite{hensgens2017a,Mukhopadhyay2018},  
crystal structures such as 1T-$\rm TaS_2$~\cite{Manzke1989,law2017a,avigo2020a},
or artificial lattices~\cite{byrnes2007a}.
However, apart from the first example (optical lattices), these systems are never perfectly 
isolated, but more or less strongly coupled to fermionic reservoirs.   
In the following, we study the impact of this coupling to the environment on the relaxation 
dynamics.

\section{Lindblad Master Equation}

\begin{figure}[ht!]
    \centering
    \includegraphics[width=0.35\textwidth,clip=true]{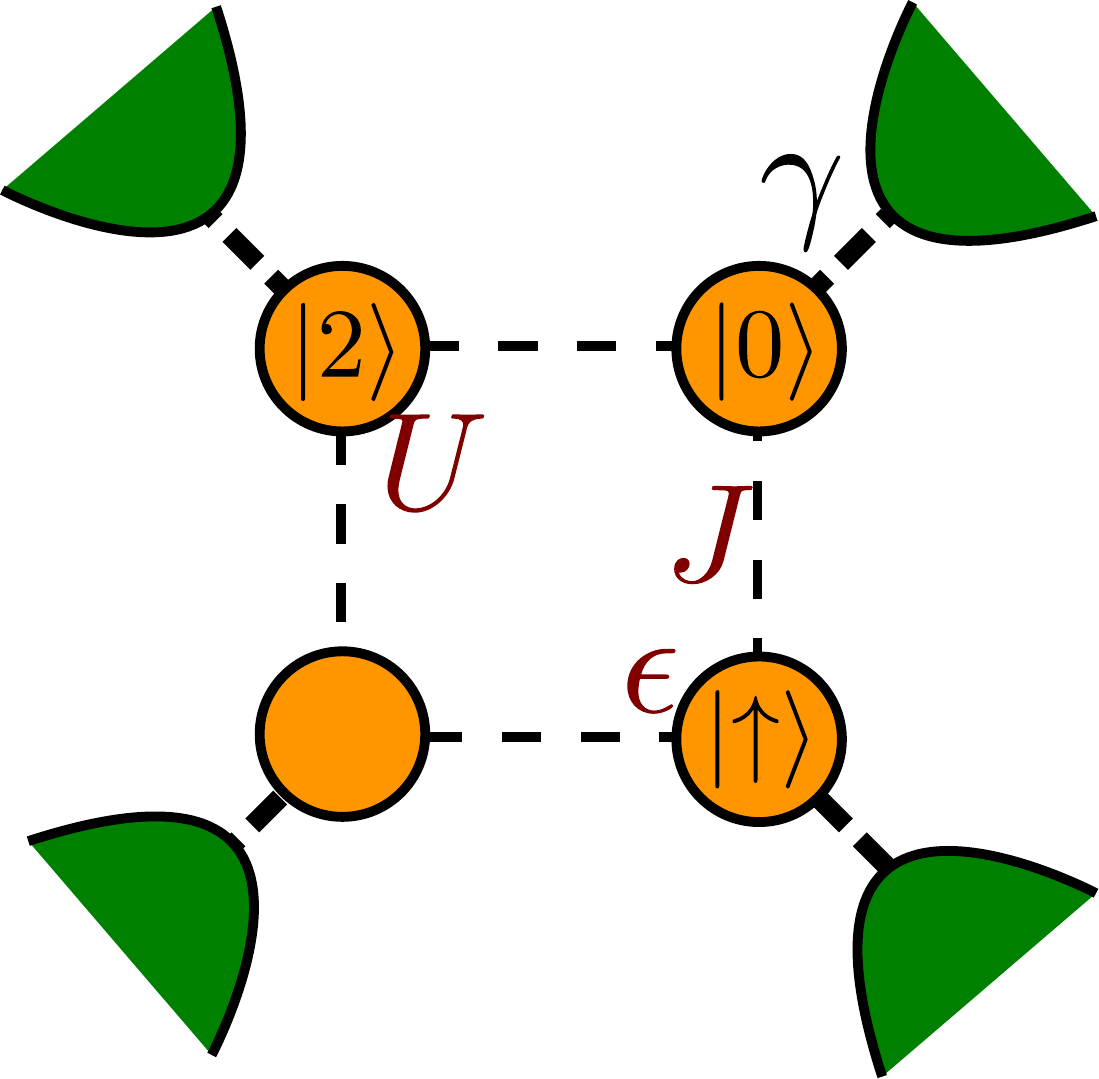}
    \caption{Sketch of a Fermi-Hubbard model~\eqref{Fermi-Hubbard} with four sites, where each site can host four states $\{\ket{0},\ket{\up},\ket{\dn},\ket{2}\}$, such that the system Hilbert space dimension is $4^4=256$. Between the sites, electronic tunnelling is possible with amplitude $J$. 
    In absence of tunnelling, a single charge on a site (bottom right) will have an on-site energy $\epsilon$, whereas two electrons of opposite spin (top left) will have the energy $2\epsilon+U$. The local reservoirs may load and unload the system with bare rate $\gamma$.}
    \label{FIG:sketch_multires}
\end{figure}

In addition to the unitary system dynamics generated by the Fermi-Hubbard
Hamiltonian~\eqref{Fermi-Hubbard}, we consider the coupling to the environment. 
Assuming that the coupling is sufficiently weak and that memory effects of the 
bath can be neglected (Born-Markov approximation), we may describe the evolution 
of the system density matrix $\hat\rho$ by a generic 
Gorini-Kossakowski-Sudarshan-Lindblad~\cite{gorini1976a,lindblad1976a} master equation ($\hbar=1$)
\begin{align}
\label{EQ:master}
\frac{d\hat\rho}{dt}
=
-\ii \left[\hat H,\hat\rho\right]
+\sum_I
\left(
\hat L_I\hat\rho\hat L_I^\dagger 
-\frac12\left\{\hat L_I^\dagger\hat L_I,\hat\rho\right\} 
\right) 
\,,
\end{align} 
where $\hat L_I$ denote the Lindblad operators 
that we motivate below.
As in~\cite{Prosen2014,carmele2015a,Popkov2015,znidaric2015a,wu2019a,queisser2019a,kleinherbers2020a}
we assume that each lattice site $\mu$ features its local set of 
Lindblad operators~\footnote{Deviations from this assumption and the consequences 
of a global bath in comparison to local reservoirs will be discussed in a 
forthcoming publication.} as is sketched in Fig.~\ref{FIG:sketch_multires}.

Let us briefly discuss the region of applicability of this description.
If we first consider disconnected lattice sites (i.e., $J=0$) weakly coupled to separate 
Markovian baths, 
such a master equation~\eqref{EQ:master} can be derived in a standard way~\cite{breuer2002}
by considering sites and reservoirs separately.  
Now switching on the tunneling $J$ between lattice sites, an analogous derivation can be 
carried out, as we show in App.~\ref{APP:local_derivation}, as long as the coupling $J$
between the lattice sites is weak in comparison to their coupling $\gamma$ to the reservoir 
-- while both are much smaller than the interaction strength $U$.

\new{
The applicability of the underlying assumptions of separate or individual baths for each 
lattice site, as well as $J\ll\gamma$, depends on the specific physical realization. 
For ultra-cold atoms in optical lattices~\cite{hofstetter2002a}, 
the Mott-Hubbard system could be realized 
in a planar lattice while the separate baths correspond to atoms moving (guided by lasers)
perpendicular to this plane -- where the in-plane hopping strength $J$ can be tuned to be 
much smaller than the perpendicular tunneling strength which determines $\gamma$. 
For Fermi-Hubbard simulators based on gate-defined quantum dots~\cite{hensgens2017a}, 
local tunnelling reservoirs can be represented by smaller tunnel electrodes placed nearby.
For other Hubbard systems sharing a common reservoir, 
the description based on separate baths can be a good approximation 
when the coupling to the lattice sites 
is effectively incoherent
(e.g. if the lattice spacing of the Hubbard system is much larger than the relevant length scale, 
such as coherence or correlation length of the reservoir;
or the bath relaxation is fast enough.) 
}

\subsection{Zero-temperature bath}

As explained above,
consistent with our assumption of a strongly interacting system, we assume that 
the on-site repulsion $U$ is not only much stronger than the system hopping $J$, 
but also dominant in comparison with the coupling to the bath. 
Apart from the spectral density of the reservoir, the impact of this coupling 
to the bath is mainly determined by the inverse temperature $\beta$ and the chemical 
potential $\mu_{\rm b}$ of the reservoir (which we assume to consist of free fermions). 
Let us first consider the case of zero temperature (cold bath), finite temperatures 
will be discussed in Sec~\ref{SEC:finite_temp} below. 

If the chemical potential of the bath is too low $\mu_{\rm b}<\epsilon$, 
all the particles from the system will 
just tunnel to the bath such that we are left with an empty system state.  
In the other limiting case, if the chemical potential is too high $\mu_{\rm b}>U+\epsilon$, the reservoir will 
fill up the system, also leading to a trivial state. 
Thus, we assume an intermediate chemical potential, for example 
$\mu_{\rm b}=\epsilon+U/2$, where only a doubly occupied site 
can release a particle into the reservoir due to its high on-site repulsion energy $U$,
while singly occupied sites cannot do that 
as the reservoir states at the energy $\epsilon$ are already filled (Pauli principle). 
Conversely, only an empty lattice site can receive a particle from the bath. 
Altogether, a cold fermionic bath at intermediate chemical potential is described by two 
Lindblad operators for each site $\mu$ and spin $s$ 
\begin{align}
\label{cold-bath}
\hat{L}_I \in 
\left\{
\sqrt{\gamma}\,
\left(1-\hat n_\mu^{\bar s}\right) 
\hat c_{\mu,s}^\dagger \,,
\sqrt{\gamma}\,
\hat n_\mu^{\bar s}
\hat c_{\mu,s} 
\right\}\,,
\end{align}
where the double index $I=\{\mu,s\}$ comprises sites and spins.
Here, $\bar s$ denotes the spin opposite to $s$ and $\gamma$ measures the strength of the 
coupling to the environment. 
Here, we assume that all these coupling strengths are the same, but one can also consider 
the case of different couplings $\gamma\to\gamma_I$. 

\section{Relaxation Dynamics at $T=0$}

Even though solving the full equation of motion~\eqref{EQ:master} is only possible for very small 
lattices, we may gain interesting insight by considering special observables. 
For the total particle number $\hat N$, we find a fast relaxation towards half filling 
\begin{align} 
\label{fast-relax}
\frac{d}{dt}\expval{\hat N}
=
\frac{d}{dt}\sum_{\mu,s}\langle\hat n_\mu^s\rangle
=
-2\gamma\sum_{\mu,s}\left(\langle\hat n_\mu^s\rangle-\frac12\right) 
\,.
\end{align} 

Another interesting observable is the total angular momentum~\cite{lieb1989a,steeb1993a,schumann2002a,essler2005,pavarini2015} 
\begin{align}
\hat{\f{S}}\label{EQ:spin}
=
\sum_\mu\hat{\f{S}}_\mu
=
\frac12
\sum_{\mu,s,s'}\hat c^\dagger_{\mu,s} \f{\sigma}_{s,s'} \hat c_{\mu,s'}
\end{align}
with the matrix elements of the Pauli matrices denoted by $\f{\sigma}_{s,s'}$.
Its components obey the usual spin algebra $[\hat{S}^x, \hat{S}^y]=\ii\hat{S}^z$, and one can define the usual ladder operators via $\hat{S}^\pm=\hat{S}^x\pm\ii \hat{S}^y$ that obey the commutation relations $[\hat{S}^-, \hat{S}^+]=-2 \hat{S}^z$ and $[\hat{S}^z, \hat{S}^\pm]= \pm \hat{S}^\pm$.
The unitary system dynamics generated by the Fermi-Hubbard Hamiltonian~\eqref{Fermi-Hubbard}
in absence of a bath conserves this quantity $[\hat H,\hat{\f{S}}]=0$ (and therefore also $\hat{\f{S}}^2$ is conserved for $\gamma=0$).
After coupling to the zero-temperature bath discussed above~\eqref{EQ:master}, 
we still obtain an exact conservation
law for the expectation value of the total spin $\hat{\f{S}}$
\begin{align}\label{EQ:spin_const}
\frac{d}{dt}\langle\hat{\f{S}}\rangle=0
\,.
\end{align} 
Unlike in the unitary case however, its square is not conserved 
\begin{align} 
\label{spin-growth}
\frac{d}{dt}\langle\hat{\f{S}}^2\rangle
=
\frac{3\gamma}{2}\sum_\mu 
\langle
\hat n_\mu^\uparrow\hat n_\mu^\downarrow 
+
(1-\hat n_\mu^\uparrow)(1-\hat n_\mu^\downarrow)  
\rangle
\geq0 
\,.
\end{align} 
Since the right-hand side of the above equation is non-negative, we find that 
$\langle\hat{\f{S}}^2\rangle$ always grows until a steady state $\hat\rho_\infty$ 
is reached.
Furthermore, we find that this steady state $\hat\rho_\infty$ must have exactly 
one particle per site
such that $\langle\hat n_\mu^\uparrow\hat n_\mu^\downarrow\rangle
=\langle(1-\hat n_\mu^\uparrow)(1-\hat n_\mu^\downarrow)\rangle=0$, 
i.e., the Mott insulator property is stable.  

In addition to the total angular momentum $\hat{\f{S}}$ discussed above, which can be 
defined for arbitrary lattice structures, we may introduce further relevant observables in 
bi-partite lattices.
For these, we can find a site ordering $\mu$ such that the parity $(-1)^\mu$ is always 
opposite for neighboring lattice sites, which -- in analogy to the spin -- allows us to introduce the pseudo-spin operators~\cite{yang1989a,zhang1990a} 
%
%
\begin{align}
\hat\eta &=\sum_\mu (-1)^\mu \hat c_{\mu,\uparrow}\hat c_{\mu,\downarrow}
\,,\quad 
\hat\eta_z=\frac12\left(\hat N-N_{\rm lattice}\right)\,,  
\end{align}
where $N_{\rm lattice}$ denotes the number of lattice sites. 
Analogous to the spin ladder operators, these obey the relations $[\hat\eta,\hat\eta^\dagger]=-2\hat\eta_z$, $[\hat\eta_z, \hat\eta]=-\hat\eta$, and $[\hat\eta_z, \hat\eta^\dagger]=+\hat\eta^\dagger$.
In contrast to Eq.~\eqref{EQ:spin_const}, the pseudo-spin is not conserved in presence of a bath.
The square of the total pseudo-spin is defined as 
\begin{align}\label{EQ:pseudospin}
\hat{\f{\eta}}^2
=
\frac12\left(\hat\eta\hat\eta^\dagger+\hat\eta^\dagger\hat\eta\right)
+\hat\eta_z^2
\,,
\end{align}
and commutes with $\hat H$, $\hat N$, $\hat{\f{S}}$, and $\hat{\f{S}}^2$. 
It obeys a simple evolution equation quite analogous to Eq.~\eqref{spin-growth}
\begin{align}
\label{EQ:pseudospin-decay}
    \frac{d}{dt} \expval{\hat{\f{\eta}}^2} 
    &=
    \frac{3\gamma}{2}\sum_\mu 
\langle
\hat n_\mu^\uparrow\hat n_\mu^\downarrow 
+
(1-\hat n_\mu^\uparrow)(1-\hat n_\mu^\downarrow)  
\rangle
\nn 
&\qquad-4 \gamma \expval{\hat{\f{\eta}}^2}
\,.
\end{align} 
Since the steady state $\hat\rho_\infty$ must have exactly one particle per site
$\langle\hat n_\mu^\uparrow\hat n_\mu^\downarrow\rangle=
\langle(1-\hat n_\mu^\uparrow)(1-\hat n_\mu^\downarrow)\rangle=0$, 
we see that $\expval{\hat{\f{\eta}}^2}$ must vanish in the steady state.

\subsection{Steady states}

Even without solving the full problem, we may infer some properties of the steady states~\cite{albert2014a,nigro2019a} 
$\hat\rho_\infty$. 
Because they have exactly one particle per lattice site, the Lindblad operators 
$\hat L_I\hat\rho_\infty=0$ 
and the interaction term 
$\hat H_U\hat\rho_\infty=0$ 
acting on these states vanish identically while $\hat H_\epsilon$ acts trivially. 
Thus, it suffices to consider only the action of the hopping Hamiltonian $\hat H_J$ 
which can be diagonalized easily.  
Since those stationary states $\hat\rho_\infty$  
must commute with $\hat H_J$, we may diagonalize them simultaneously. 
Then, because each term $\hat c^\dagger_{\mu,s}\hat c_{\nu,s}$ in the hopping Hamiltonian
$\hat H_J$, 
after acting on a state with exactly one particle per lattice site, either annihilates this 
state or leads to a doubly occupied ($\mu$) and an empty ($\nu$) lattice site, these steady 
states $\hat\rho_\infty$ must also be in the sub-space with zero eigenvalue of the 
hopping Hamiltonian $\hat H_J$, i.e., $\hat H_J\hat\rho_\infty=0$. 
One example is the ferromagnetic state $\ket{\uparrow\uparrow\uparrow\dots}$ which 
maximizes $\hat S_z$ and $\hat{\f{S}}^2$ and is obviously a steady state.
Now, as the ladder operators $\hat S_\pm$ commute with $\hat H$, we see that all the 
states $\ket{\Psi_n}\propto\hat S_-^n\ket{\uparrow\uparrow\uparrow\dots}$ are steady states. 
They maximize $\hat{\f{S}}^2$ and form a ladder with $N_{\rm lattice}+1$ states 
from $\ket{\uparrow\uparrow\uparrow\dots}$ 
to $\ket{\downarrow\downarrow\downarrow\dots}$, whose rungs
can be labeled by their different eigenvalues of $\hat S_z$.  
%

\subsection{Anti-ferromagnetic order}

As we found above, the states with maximum $\hat{\f{S}}^2$ are steady states. 
Note however, that the previous line of arguments does not necessarily imply that these are the 
only steady states -- this depends on the lattice structure 
(which determines the diagonalization of $\hat H_J$). 
For a lattice which can be decomposed into two disconnected sub-lattices, for example, 
there are further steady states (which maximize $\hat{\f{S}}^2$ for each sub-lattice
separately). 

However, the anti-ferromagnetic state (which is the ground state of $\hat H$ on 
simply connected bi-partite lattices) is not a steady state. 
As we have seen above, the steady states are annihilated by $\hat H_J$, i.e., all the hopping 
contributions vanish or cancel each other.
In contrast, the anti-ferromagnetic order of the Mott-N\'eel state is precisely such that it 
facilitates tunneling in order to lower the energy~\cite{cleveland1976a}.

As a result, while the Mott insulator structure itself is stable (i.e., conserved by the 
steady states), we find a decay of the anti-ferromagnetic order of the Mott-N\'eel state 
due to the coupling to the environment according to Eq.~\eqref{EQ:master}. 

\subsection{Relaxation time scales}

Finally, let us discuss the different time scales of relaxation. 
Obviously, the relaxation to half filling occurs on a time scale $\ord(1/\gamma)$.
In contrast, in the strongly interacting case $U\gg J$ at half filling, the right-hand 
side of~\eqref{spin-growth} is suppressed as $J^2/U^2$ (as already mentioned in the introduction) 
and thus the 
growth in~\eqref{spin-growth} is much slower than that in~\eqref{fast-relax}.
As a result, we find two vastly different relaxation time scales -- as already observed 
in many other systems and scenarios, see, e.g., Refs.~\cite{queisser2014a,queisser2019a,kleinherbers2020a,wu2020a,avigo2020a,wang2020a}.
A more quantitative analysis of these two time scales will be presented in the next section. 

\section{Waiting Time Distribution}

In order to study the different time scales observed above in more detail, let us re-write 
the master equation~\eqref{EQ:master} in terms of the Liouville super-operator $\cal L$
%
\begin{align}
\label{Liouville}
\frac{d\hat\rho}{dt}
=
-\ii\hat H_{\rm eff}\hat\rho+\ii\hat\rho\hat H_{\rm eff}^\dagger 
+\sum_I
\hat L_I\hat\rho\hat L_I^\dagger 
=
{\cal L}\hat\rho
\,,
\end{align} 
where we have introduced the effective Hamiltonian 
\begin{align}
\label{effective-Hamiltonian}
\hat H_{\rm eff}
=
\hat H-\frac{\ii}{2}\,\sum_I\hat L_I^\dagger\hat L_I
=
\hat H-\frac{\ii}{2}\,\hat{\f{L}}^2 
\,,
\end{align} 
which is non-Hermitian due to the coupling with the bath.
For the cold bath~\eqref{cold-bath}, we find 
\begin{align}
\hat{\f{L}}^2 
=
-2\gamma\sum_{\mu,s}\left(\hat n_\mu^s -\frac12\right)
+4\gamma\sum_\mu \hat n_\mu^\uparrow\hat n_\mu^\downarrow 
\,.
\end{align}
Thus, apart from the c-number contribution, the effective 
Hamiltonian~\eqref{effective-Hamiltonian} has the same structure as the 
Fermi-Hubbard model~\eqref{Fermi-Hubbard}, just with complex parameters 
$U\to U-2\ii\gamma$ and $\epsilon\to\epsilon+\ii\gamma$. 

As the next step, we decompose the total Liouville super-operator $\cal L$ 
into the jump term 
\begin{align}
{\cal L}_{\rm jump}\hat\rho
=
\sum_I\hat L_I\hat\rho\hat L_I^\dagger 
\end{align} 
and remaining contribution ${\cal L}_{\rm stay}={\cal L}-{\cal L}_{\rm jump}$
corresponding to no jumps.  
Then, the total evolution of the density matrix $\hat\rho$ in a time interval $[0,t]$ 
can be expanded into a Dyson series in ${\cal L}_{\rm jump}$ via 
\begin{align}
e^{{\cal L}t}
=
e^{{\cal L}_{\rm stay}t}
+\int_0^t dt'\,e^{{\cal L}_{\rm stay}(t-t')} {\cal L}_{\rm jump} e^{{\cal L}_{\rm stay}t'}
+\dots 
\end{align}
The first term $e^{{\cal L}_{\rm stay}t}$ corresponds to no jumps in time interval $[0,t]$, 
the second term describes one jump at a time $t'$ and the higher-order contributions 
correspond to two or more jumps.  
As a result, starting with the initial state $\hat\rho_0=\hat\rho(t=0)$, 
the probability for no jump during the time interval $[0,t]$ is given by 
\begin{align}
P_{\rm stay}(t)
=
{\rm Tr}\left\{e^{{\cal L}_{\rm stay}t}\hat\rho_0\right\} 
=
{\rm Tr}\left\{e^{-\ii\hat H_{\rm eff}t}
\hat\rho_0 e^{+\ii\hat H_{\rm eff}^\dagger t}
\right\}.
\end{align}
This allows us to infer the waiting time distribution~\cite{cohen_tannouji1986a,plenio1998a,brandes2008a,albert2011a,albert2012a,rajabi2013a,sothmann2014a,ptaszynski2017a,kleinherbers2021a},
i.e., the probability density 
$\omega(t)$ of the waiting time $t$ until an initial state $\hat\rho_0$ decays due to the first jump, via 
\begin{align}
\omega(t)=-\dot{P}_{\rm stay}\,.
\end{align}
This waiting time distribution is a quantitative measure of the relaxation rates and 
its decay constants are determined by the complex eigenvalues $\sigma_n$ of $\hat{H}_{\rm eff}$
via $e^{-\ii H_{\rm eff} t}=\sum_n {\hat M}_n(t) e^{-\ii \sigma_n t}$, where the ${\hat M}_n(t)$ can (for degenerate $\sigma_n$) be polynomial functions of $t$.
Thus, the imaginary parts $\Im{\sigma_n}\le 0$ in particular determine the decay characteristics of the waiting time distributions.

A first rough estimate of these eigenvalues of $\hat H_{\rm eff}$ can be obtained by 
perturbation theory in $\gamma$. 
Note that perturbation theory for non-Hermitian operators (such as $\hat H_{\rm eff}$) 
is typically more complicated than for Hermitian operators (such as $\hat H$)~\cite{buth2004a,Ashida2020}.
Nevertheless, starting from the undisturbed eigenstates $\ket{u_\lambda}$ of the 
Fermi-Hubbard Hamiltonian $\hat H$ corresponding to the real undisturbed 
(and assumed to be non-degenerate) eigenvalues $\lambda$, we may find the first-order 
shift of the associated complex eigenvalues $\sigma_\lambda$ of $\hat H_{\rm eff}$ by 
\begin{align}
\sigma_\lambda=\lambda-\frac{\ii}{2}\bra{u_\lambda}\hat{\f{L}}^2\ket{u_\lambda}
+\ord(\gamma^2)
\,.
\end{align}
For example, the empty state $\ket{u_\lambda}=\ket{00\dots}$ with $\lambda=0$ 
is shifted to the imaginary eigenvalue $\sigma_\lambda=-\ii\gamma N_{\rm lattice}$ 
with $N_{\rm lattice}$ denoting the number of lattice sites -- which corresponds to the 
fast relaxation of the particle number~\eqref{fast-relax}. 
As another example, the aforementioned ferromagnetic state 
$\ket{u_\lambda}=\ket{\uparrow\uparrow\dots}$ 
is inert and thus has no imaginary shift $\Im(\sigma_\lambda)=0$, 
showing that it is a steady state (at zero temperature). 

The evaluation of waiting times between two jumps is also 
possible~\cite{brandes2008a,albert2011a}, and can analogously be achieved by evaluating 
the exponential of $\hat{H}_{\rm eff}$.

\section{Hubbard Tetramer at $T=0$}

The above approach based on the waiting-time distribution goes along with a tremendous 
reduction in complexity.
Instead of calculating $e^{{\cal L}t}$ or diagonalizing $\cal L$, it suffices to 
diagonalize the effective Hamiltonian $\hat H_{\rm eff}$.
Moreover, we found that $\hat H_{\rm eff}$ has the same structure as the original 
Fermi-Hubbard Hamiltonian $\hat H$, just with complex parameters $U$ and $\epsilon$.
Of course, despite this reduction in complexity, one can only derive general analytic 
expressions in those cases where the original Fermi-Hubbard Hamiltonian $\hat H$ 
can be diagonalized. 

In order to treat such a simple (yet non-trivial) case, we consider the Fermi-Hubbard 
model on a square (which is equivalent to a ring consisting of 4 lattice sites). 
Even in this simple case, the total Hilbert space contains $4^4=256$ states,
i.e., $\hat H_{\rm eff}$ and $\hat H$ can be represented as $256\times256$-matrices. 
In order to bring these matrices into a treatable block-diagonal form, we employ 
a suitable set of commuting observables.   

Apart from the total particle number $\hat N$, we select $\hat S_z$ and $\hat{\f{S}}^2$ 
as further commuting observables. 
The total spin $\hat{\f{S}}^2$ allows us to classify the Hubbard tetramer spectrum into 
42 singlets, 48 doublets, 27 triplets, 8 quadruplets, and 1 quintuplet. 
These states compose the 
$42 \cdot 1 + 48 \cdot 2 + 27 \cdot 3 + 8 \cdot 4 + 1 \cdot 5 = 256$ states in the 
total Hilbert space.  
In analogy to the spin $\hat{\f{S}}^2$, the pseudo-spin~\eqref{EQ:pseudospin} 
also allows to classify the 256 states 
into $42$ pseudo-singlets, $48$ pseudo-doublets, $27$ pseudo-triplets, $8$ pseudo-quadruplets, 
and $1$ pseudo-quintuplet.
The classification according to the pseudo-spin is different from that according to the spin, 
which allows us to decompose the Hilbert space further.

As the final ingredient, we use suitable geometric symmetries of the square. 
One option could be the quasi-momentum 
which generates the cyclic permutation $\hat{P}_{1234}$:
$1\to2$, $2\to3$, $3\to4$ and $4\to1$ of the lattice sites, see App.~\ref{APP:quasimom}.
However, we found it more convenient to employ the reflections at the two diagonals 
of the square, i.e., 
the permutations $\hat P_{13}$ and $\hat P_{24}$
\begin{align}
\hat P_{\mu\nu}
=
\exp\left\{
\ii \frac{\pi}{2}\sum_s\left(
\hat c_{\mu,s}^\dagger\hat c_{\nu,s}+ 
\hat c_{\nu,s}^\dagger\hat c_{\mu,s}-
\hat n_{\mu,s}-\hat n_{\nu,s}
\right)  
\right\} 
\,,
\nn
\end{align} 
exchanging sites $1\leftrightarrow3$ and $2\leftrightarrow4$, respectively.
These symmetry operations represent additional 
commuting observables with eigenvalues (parities) $\pm1$. 

Now we may use the set 
$\{\hat N,\hat S_z,\hat{\f{S}}^2,\hat{\f{\eta}}^2,\hat P_{13},\hat P_{24}\}$ 
to diagonalize $\hat H_{\rm eff}$, which can be decomposed into independent blocks 
with maximum rank four, which allows us to find the eigenvalues analytically.
For example, the empty state $\ket{0000}$ lies in the sector with even parities 
where all the other quantities 
$\{\hat N,\hat S_z,\hat{\f{S}}^2,\hat{\f{\eta}}^2\}$ vanish.
It is an eigenstate of $\hat H_{\rm eff}$ with the eigenvalue $\sigma=-4\ii\gamma$, and its waiting-time distribution correspondingly reads 
\begin{align}
w(t)=-\dot P_{\rm stay}(t)=8\gamma e^{-8\gamma t}
\,.
\end{align} 
As already explained above, the quintuplet states are steady states %
\begin{align}
\label{EQ:quintuplett}
\ket{\Psi_{-2}} &= \ket{\dn\dn\dn\dn}\,,\nn
\ket{\Psi_{-1}} &= \frac{1}{2}\left[\ket{\up\dn\dn\dn}+\ket{\dn\up\dn\dn}+\ket{\dn\dn\up\dn}+\ket{\dn\dn\dn\up}\right]\,,\nn
\ket{\Psi_{0}}  &=\frac{1}{\sqrt{6}} 
\Big[\ket{\up\up\dn\dn}+\ket{\up\dn\up\dn}+\ket{\up\dn\dn\up}\nn
&\qquad+\ket{\dn\up\up\dn}+\ket{\dn\up\dn\up}+\ket{\dn\dn\up\up}\Big]
\,,\nn
\ket{\Psi_{+1}} &= \frac{1}{2}\left[\ket{\dn\up\up\up}+\ket{\up\dn\up\up}+\ket{\up\up\dn\up}+\ket{\up\up\up\dn}\right]\,,\nn
\ket{\Psi_{+2}} &= \ket{\up\up\up\up}\,,
\end{align}
and thus their waiting-time distributions vanish identically $w(t)=0$.
They all contain 
four particles with maximum $\hat{\f{S}}^2$ and are labeled by 
their $\hat S_z$ eigenvalues.
All these states have odd parities (due to the Pauli principle)
and are annihilated by $\hat{\f{\eta}}^2$, i.e., $\abs{\eta}=0$.
For the Hubbard tetramer, these are the only steady states. 

For the Hubbard dimer, we found in Ref.~\cite{kleinherbers2020a} a slow relaxation for the 
spin singlet state $(\ket{\uparrow\downarrow}-\ket{\downarrow\uparrow})/\sqrt{2}$.
Thus, let us now consider its straight-forward generalization to four lattice sites 
\begin{align}
\label{EQ:af_as}
\ket{\Psi_{\rm zaf}}=
\frac{\ket{\uparrow\downarrow\uparrow\downarrow}-\ket{\downarrow\uparrow\downarrow\uparrow}}
{\sqrt{2}}
\,.
\end{align} 
This state displays Ising type (i.e., $\hat S_z$) anti-ferromagnetic order and lies  
in the sector with $N=4$ particles, $\abs{S}=1$, 
$S_z=0$, and $\abs{\eta}=0$ and is separately odd under the site permutations 
$\hat{P}_{13}$ and $\hat{P}_{24}$. 
This sector contains~\footnote{One actually finds in total four states with $N=4$ particles, $\abs{S}=1$, 
$S_z=0$, and $\abs{\eta}=0$ and odd parities $P_{13}=-1=P_{24}$, but one of these can be 
trivially decoupled.
If one classifies the states by the more complicated quasi-momentum instead, the sector of 
interest contains only the three states discussed.}
two additional states $\ket{\Psi_{\rm zaf}^1}$ and  
$\ket{\Psi_{\rm zaf}^2}$ that can be 
conveniently generated from $\ket{\Psi_{\rm zaf}}$ by acting with the hopping Hamiltonian 
$\hat H_J$ and subsequent Erhard-Schmidt orthogonalization~\footnote{\label{FN:note1}
Acting with the hopping Hamiltonian on the state $\ket{\Psi^0}\in\{\ket{\Psi_{\rm zaf}},\ket{\Psi_{\rm af}}\}$, we obtain the already orthogonal state
$\ket{\Psi^1} \propto \hat H_J \ket{\Psi^0}$.
Acting again with the hopping Hamiltonian and orthogonalizing we obtain
$\ket{\Psi^2} \propto \hat H_J \ket{\Psi^1}
-(\bra{\Psi^0} \hat H_J \ket{\Psi^1}) \ket{\Psi^0}$.}.

In the basis $\{\ket{\Psi_{\rm zaf}^0}=\ket{\Psi_{\rm zaf}},\ket{\Psi_{\rm zaf}^1},\ket{\Psi_{\rm zaf}^2}\}$,
the effective Hamiltonian can be represented by the $3\times3$-matrix
(the signs of off-diagonal matrix elements can be controlled by properly choosing 
the basis vectors)
\begin{align}\label{EQ:heff_zaf_zerotemp}
\hat H_{\rm eff} 
=
\left(\begin{array}{ccc}
4\epsilon & +\sqrt{8} J & 0\\
+\sqrt{8} J & 4\epsilon+U-2\ii\gamma & +\sqrt{8} J\\
0 & +\sqrt{8} J & 4\epsilon+U-2\ii\gamma
\end{array}\right)\,.
\end{align}
The eigenvalues of this matrix reflect the two relaxation times scales mentioned above. 
In the strongly interacting limit $U\gg J$, two eigenvalues $\sigma_\pm$ have a large 
imaginary part $\Im(\sigma_\pm)\approx-2\gamma$ while the remaining eigenvalue $\sigma_0$ 
-- whose eigenvector has the largest overlap with the anti-ferromagnetic 
state~\eqref{EQ:af_as}
-- has much smaller imaginary part $\Im(\sigma_0)=\ord(\gamma J^2/U^2)$.  
Specifically, a Taylor expansion in $J/U\ll1$ up to quadratic order yields 
\begin{align}\label{EQ:evals_heff_zaf}
\sigma_0 &\approx 4\epsilon -\frac{8 J^2}{U} -\ii \frac{16 \gamma J^2}{U^2}\,,\\
\sigma_\pm &\approx 4\epsilon+U\pm2\sqrt{2}J+\frac{4J^2}{U} 
+\ii \left(-2\gamma+\frac{8\gamma J^2}{U^2}\right)\,.
\nonumber
\end{align}

Finally, let us consider the full Heisenberg type anti-ferromagnetic state, 
i.e., the Mott-N\'eel state
\begin{align}\label{EQ:af_gs}
    \ket{\Psi_{\rm af}} &= \frac{1}{\sqrt{12}}\big[2\ket{\up\dn\up\dn}+2\ket{\dn\up\dn\up}\nn
    &\qquad-\ket{\up\up\dn\dn}-\ket{\up\dn\dn\up}-\ket{\dn\dn\up\up}-\ket{\dn\up\up\dn}\big]
    \,,
\end{align}
which is orthogonal to the previously discussed states~\eqref{EQ:quintuplett} and~\eqref{EQ:af_as}. 
It belongs to the sector with $N=4$, $S_z=0$, $\abs{S}=0$, $\abs{\eta}=0$, and is separately odd under the site permutations $\hat{P}_{13}$ and $\hat{P}_{24}$.
As before, this sector contains two additional states that can be created by the action of $\hat{H}_J$ and subsequent orthonormalization~\footnotemark[2], and in the basis formed by $\{\ket{\Psi_{\rm af}^0}=\ket{\Psi_{\rm af}}, \ket{\Psi_{\rm af}^1}, 
\ket{\Psi_{\rm af}^2}\}$,
the effective non-Hermitian Hamiltonian has the representation
\begin{align}\label{EQ:heff_totaf_zerotemp}
    \hat{H}_{\rm eff} &= \left(\begin{array}{ccc}
    4\epsilon & \sqrt{12} J & 0\\
    \sqrt{12} J & 4\epsilon+U-2\ii\gamma & 2 J\\
    0 & 2 J & 4\epsilon+2U-4\ii\gamma 
    \end{array}\right)\,.
\end{align}
Taylor expansion for $J/U \ll 1$ yields the eigenvalues
\begin{align}\label{EQ:evals_heff_totaf}
    \sigma_0 &\approx 4\epsilon-12 \frac{J^2}{U}-\ii \frac{24\gamma J^2}{U^2}\,,\nn
    \sigma_1 &\approx 4\epsilon+U+ 8 \frac{J^2}{U} +\ii \left(-2\gamma + \frac{16 \gamma J^2}{U^2}\right)\,,\nn
    \sigma_2 &\approx 4\epsilon+2U+ 4 \frac{J^2}{U} +\ii \left(-4\gamma + \frac{8 \gamma J^2}{U^2}\right)\,,
\end{align}
which now has three distinct modes, two fast decaying ones and a slower decaying one.

Retrieving the isolated Fermi-Hubbard model by considering the real part of Eqns.~\eqref{EQ:evals_heff_zaf} and~\eqref{EQ:evals_heff_totaf}, 
we see that the lowest energy 
state comes from the sector described by matrix representation~\eqref{EQ:heff_totaf_zerotemp}, 
and indeed one can show that the ground state in the sector with $N_{\up}=N_{\dn}=2$ is spanned 
by the states $\ket{\Psi_{\rm af}^0}$ and (for $J\ll U$ small) 
contributions of $\ket{\Psi_{\rm af}^1}$ and $\ket{\Psi_{\rm af}^2}$.
Indeed, it is well-known~\cite{lieb1989a} that the ground state of the half-filled sector must 
have $\langle\hat{\f{S}}^2\rangle=0$.
Given the evident anti-ferromagnetic order of~\eqref{EQ:af_as}, this may appear surprising 
but one should keep in mind that the true ground state maximizes 
the full (Heisenberg type) anti-ferromagnetic order operator 
\begin{align}
    \hat{O}_{\rm af} = -\sum_{\mu=1}^4 \hat{\f{S}}_\mu \cdot \hat{\f{S}}_{\mu+1}\,,
\end{align}
which is directly proportional to a Heisenberg Hamiltonian.
This order operator commutes with $\hat{N}$, $\hat{S}_z$, $\hat{\f{S}}^2$, and the 
quasi-momentum $\hat{P}_{1234}$ or the permutation operators $\hat{P}_{13}$ and $\hat{P}_{24}$, 
it can therefore also be brought in the same block-diagonal form as $\hat{H}_{\rm eff}$.
Direct inspection of $\matel{\Psi_{\rm af}}{\hat{O}_{\rm af}}{\Psi_{\rm af}}=2$ shows that $\ket{\Psi_{\rm af}}$ is maximally ordered, and thereby more ordered than
$\ket{\Psi_{\rm zaf}}$, for which one finds $\matel{\Psi_{\rm zaf}}{\hat{O}_{\rm af}}{\Psi_{\rm zaf}}=1$.
Even for finite but small $J/U$, the ground states (for $\gamma=0$) of Eqns.~\eqref{EQ:heff_zaf_zerotemp} and~\eqref{EQ:heff_totaf_zerotemp} possess the order parameters
$\langle\hat{O}_{\rm af}\rangle\approx 1-10 J^2/U^2$ and $\langle\hat{O}_{\rm af}\rangle\approx 2-15 J^2/U^2$, respectively.
One can show that these expectation values are consistent with the lowest eigenvalues~\eqref{EQ:evals_heff_zaf} and~\eqref{EQ:evals_heff_totaf} of the isolated tetramer, for which an effective Heisenberg-type Hamiltonian applies~\cite{cleveland1976a}, see App.~\ref{APP:heisenberg}.

In an analogous fashion, the decay properties of all eigenstates of the Hubbard model on the 
square can be analytically evaluated.
For illustration, we provide the qualitative decay dynamics of all $36$ states in the 
sector $N_\up=2=N_\dn$ in App.~\ref{APP:decay_chart}.

\section{Finite-temperature bath}\label{SEC:finite_temp}

For finite reservoir temperatures, the calculations are essentially analogous, we just have 
to take the finite occupations of the reservoir into account.
Then, the Lindblad operators in~\eqref{cold-bath} generalize to four operators per site and spin 
\begin{align}\label{EQ:ft-bath}
\hat{L}_I &\in\Big\{\sqrt{\gamma f_{\rm E}} (1-\hat{n}_\mu^{\bar s}) \hat{c}_{\mu, s}^\dagger, \sqrt{\gamma (1-f_{\rm E})} (1-\hat{n}_\mu^{\bar s}) \hat{c}_{\mu,s}\,,\nn
&\qquad\sqrt{\gamma f_{\rm U}} \hat{n}_\mu^{\bar s} \hat{c}_{\mu, s}^\dagger,\sqrt{\gamma(1-f_{\rm U})} \hat{n}_\mu^{\bar s} \hat{c}_{\mu,s}\Big\}\,,
\end{align}
\new{which we motivate microscopically in App.~\ref{APP:markseclocal}.}
Here, the thermal properties of the reservoir are encoded in the Fermi functions for transitions between empty (E) and singly-charged dots
$f_{\rm E}=[e^{\beta(\epsilon-\mu_{\rm b})}+1]^{-1}$ and for transitions between singly and doubly charged (U) dots $f_{\rm U}=[e^{\beta(\epsilon+U-\mu_{\rm b})}+1]^{-1}$,
with inverse temperature $\beta$ and chemical potential $\mu_{\rm b}$ while 
$\gamma$ denotes the spectral density 
(assumed flat over the energy scales of the tetramer, wideband limit).
The first Lindblad operator above describes a transfer
of an electron with spin $s$ onto site $\mu$, 
provided the other spin species $\bar s$ is not present at that site and the second term 
describes the reverse process, 
i.e., an electron with spin $s$ leaving site $\mu$ when the other spin species is not present.
The second line above describes the same processes when the other spin species is present, 
where the corresponding transition rates are modified by the Coulomb interaction.
When $\mu_{\rm b}=\epsilon+U/2$ and $\beta U \gg 1$, we get $f_{\rm E}\to 1$ and $f_{\rm U}\to 0$, such that the previous Lindblad
operators~\eqref{cold-bath} are reproduced.
The generator~\eqref{EQ:master} with Lindblad
operators~\eqref{EQ:ft-bath} tends to locally 
thermalize the system, i.e., in the limit $J=0$ the state 
$\hat\rho_\beta \propto \exp\left\{-\beta (\hat{H}_\epsilon+\hat{H}_U - \mu_{\rm b} \hat{N})\right\}$ 
is a stationary state.

To identify the non-Hermitian Hamiltonian, we evaluate
\begin{align}\label{EQ:jump_ops1}
\hat{\f{L}}^2 
&= 
\sum_{\mu s} \Big[\gamma f_{\rm E} (\f{1}-\hat{n}_\mu^{\bar s}) (\f{1}-\hat{n}_\mu^s) + \gamma(1-f_{\rm E}) (\f{1}-\hat{n}_\mu^{\bar s}) \hat{n}_\mu^s\nn
&\qquad+\gamma f_{\rm U} \hat{n}_\mu^{\bar s} (\f{1}-\hat{n}_\mu^s) + \gamma (1-f_{\rm U}) \hat{n}_\mu^{\bar s} \hat{n}_\mu^s\Big]\nn
&= 2 N_{\rm lattice} \gamma f_{\rm E} 
\f{1} +\gamma(1+f_{\rm U}-3f_{\rm E}) \sum_{\mu s} \hat{n}_\mu^s\nn
&\qquad+ 4\gamma (f_{\rm E}-f_{\rm U}) \sum_{\mu s} \hat{n}_\mu^{\bar s} \hat{n}_\mu^s\,.
\end{align}
Thus, apart from a shift of $-N_{\rm lattice}\ii\gamma f_{\rm E}$, the effective non-Hermitian 
Hamiltonian contains the on-site energy $\epsilon\to \epsilon-\ii\gamma/2(1+f_{\rm U}-3 f_{\rm E})$ 
and the Coulomb interaction $U\to U-2\ii\gamma (f_{\rm E}-f_{\rm U})$ as complex parameters also for 
finite temperatures.
The previously discussed zero-temperature limit (when $\mu_{\rm b}=\epsilon+U/2$ and $\beta U \gg 1$) 
is recovered by $f_{\rm E}\to 1$ and $f_{\rm U}\to 0$.

\subsection{Infinite temperature limit}

At infinite temperatures, we just set 
$f_{\rm E}\to1/2$ and $f_{\rm U}\to 1/2$, and the effective non-Hermitian Hamiltonian is trivially shifted
$\hat{H}_{\rm eff} = \hat{H}-\frac{\ii}{2}\gamma N_{\rm lattice}\f{1}$. 
Explicit evaluation of the master equation~\eqref{EQ:master} with Lindblad operators~\eqref{EQ:ft-bath} then shows that $\hat\rho_\infty\propto \f{1}$ 
is a stationary state at infinite temperatures.
Consistently, we find that in this limit observables that are conserved for the isolated Fermi-Hubbard model obey simple closed equations of motion.
For example, the particle number again decays quickly towards half filling, 
but the relaxation rate is only half that of the low-temperature limit~\eqref{fast-relax}
\begin{align}
    \frac{d}{dt} \sum_{\mu,s} \expval{\hat{n}_\mu^s} = -\gamma \left(\sum_{\mu,s} \expval{\hat{n}_\mu^s}-\frac{1}{2}\right)\,.
\end{align}
%
The spin components are no longer conserved as in~\eqref{EQ:spin_const} but decay according to
\begin{align}
    \frac{d}{dt} \expval{\hat{\f{S}}} = -\gamma \expval{\hat{\f{S}}}\,.
\end{align}
Analogously, in contrast to~\eqref{spin-growth}, the spin quadrature decays according to
\begin{align}
    \frac{d}{dt}\expval{\hat{\f{S}}^2} = -2\gamma \left(\expval{\hat{\f{S}}^2}-\frac{3}{2}\right)\,.
\end{align}
The stationary value $3/2$
of the spin quadrature is consistent with $\hat{\rho}_\infty\propto \f{1}$.

Furthermore, in contrast to~\eqref{EQ:pseudospin-decay}, the pseudo-spin evolves similarly to 
the spin
\begin{align}
     \frac{d}{dt}\expval{\hat{\f{\eta}}^2} = -2\gamma \left(\expval{\hat{\f{\eta}}^2}-\frac{3}{2}\right)\,.
\end{align}

One may ask how the interacting Lindblad operators~\eqref{EQ:ft-bath} are compatible with the 
following phenomenological picture~\cite{kleinherbers2020a}: 
For infinite temperatures, 
due to availability of the full range of energies,
one might expect that the model behaves as independent sites which 
can be loaded and unloaded just as non-interacting electrons, i.e., independent of whether an electron is already present or not. 
Closer inspection shows that for states $\hat\rho$ obeying $\left[\hat\rho,\hat n_\mu^s\right]=0$ 
(e.g., having definite particle numbers $\hat n_\mu^s$ for each site $\mu$ and spin $s$),
one may recombine the dissipators~\eqref{EQ:ft-bath} 
to recover the local description of independent electrons -- formally equivalent to Eq.~\eqref{EQ:noninteracting} for $f=1/2$,  
which would also result from the \textit{coherent approximation} introduced in Ref.~\cite{kleinherbers2020a}.
This does of course depend on the initial state, and it is indeed an interesting route of further research to investigate how the decay dynamics depends on the presence of correlations.

\subsection{Hubbard tetramer}

For finite temperatures and finite chemical potentials the initially empty ($E$) and initially 
filled ($F$) states decay trivially
\begin{align}\label{EQ:wtdef}
\omega_E(\tau) &= 8\gamma f_{\rm E} e^{-8 \gamma\tau f_{\rm E}}\,,\nn
\omega_F(\tau) &= 8\gamma (1-f_{\rm U}) e^{-8\gamma\tau(1-f_{\rm U})}\,.
\end{align}
In general, when we consider states that are completely determined by the known quantum numbers, 
the effective non-Hermitian Hamiltonian will be a $1\times 1$-matrix, which only allows the 
waiting-time distribution to decay in a simple exponential fashion. 
For example, the five quintuplett states~\eqref{EQ:quintuplett} that are completely determined 
by fixing $\hat{\f{S}}^2$ and $\hat{S}_z$
and superpositions of them all obey the same waiting time distribution
\begin{align}\label{EQ:wtds2}
\omega_{2}(\tau) = 4\gamma(1-f_{\rm E}+f_{\rm U}) e^{-4\gamma\tau(1-f_{\rm E}+f_{\rm U})}\,.
\end{align}
As mentioned, for $\mu_{\rm b}=\epsilon+U/2$, these states remain stable in the zero-temperature 
limit $\beta U \gg 1$.
For small but finite temperatures, we find that
$1-f_{\rm E}+f_{\rm U} \approx 2 e^{-\beta U/2}$, such that their average lifetime scales as
\begin{align}
\expval{\tau} = \int_0^\infty \tau \omega_2(\tau) d\tau \approx 
\frac{e^{U \beta/2}}{8 \gamma}
\,.
\end{align}
One could imagine to exploit this very long lifetime for storing (quantum) information 
in superpositions of these states at low temperatures.
However, as the waiting time distribution~\eqref{EQ:wtds2} is a simple exponential decay, 
it offers no lifetime guarantee, i.e., arbitrarily short lifetimes are possible 
(and actually always more probable than long ones).

Note that this is different for the state~\eqref{EQ:af_as}, 
for which the effective non-Hermitian Hamiltonian 
assumes nearly the same form as for zero temperatures, the only difference to Eq.~\eqref{EQ:heff_zaf_zerotemp} is that the top-left matrix element is modified to
\begin{align}
\bra{\Psi_{\rm zaf}^0} \hat{H}_{\rm eff} \ket{\Psi_{\rm zaf}^0} = 
4\epsilon-2\ii\gamma(1+f_{\rm U}-f_{\rm E})\,,
\end{align}
where we again reproduce the previous case at low temperatures and half-filling potential 
($f_{\rm E}\to 1$ and $f_{\rm U}\to 0$).
The situation is however fundamentally different in the infinite temperature limit 
($f_{\rm E}=f_{\rm U}=1/2$), where the eigenvalues of $H_{\rm eff}$ are just the eigenvalues of 
$\hat{H}$ shifted by
$-2\ii\gamma$ towards the lower complex plane.
Thus, at high temperatures this state will decay in a trivial exponential fashion $\propto e^{-4\gamma t}$, 
whereas at low temperatures the competing much slower timescale from~\eqref{EQ:evals_heff_zaf} 
is revealed.

For the fully anti-ferromagnetic state~\eqref{EQ:af_gs}, we find that in Eq.~\eqref{EQ:heff_totaf_zerotemp} two matrix elements have to be modified
\begin{align}
    \bra{\Psi_{\rm af}^0} \hat{H}_{\rm eff} \ket{\Psi_{\rm af}^0} &= 4\epsilon-2\ii\gamma(1+f_{\rm U}-f_{\rm E})\,,\nn
    \bra{\Psi_{\rm af}^2} \hat{H}_{\rm eff} \ket{\Psi_{\rm af}^2} &= 4\epsilon+2U-2\ii\gamma(1+f_{\rm E}-f_{\rm U})\,,    
\end{align}
such that also here at high temperatures all modes decay similarly.

\begin{figure}
\includegraphics[width=0.45\textwidth,clip=true]{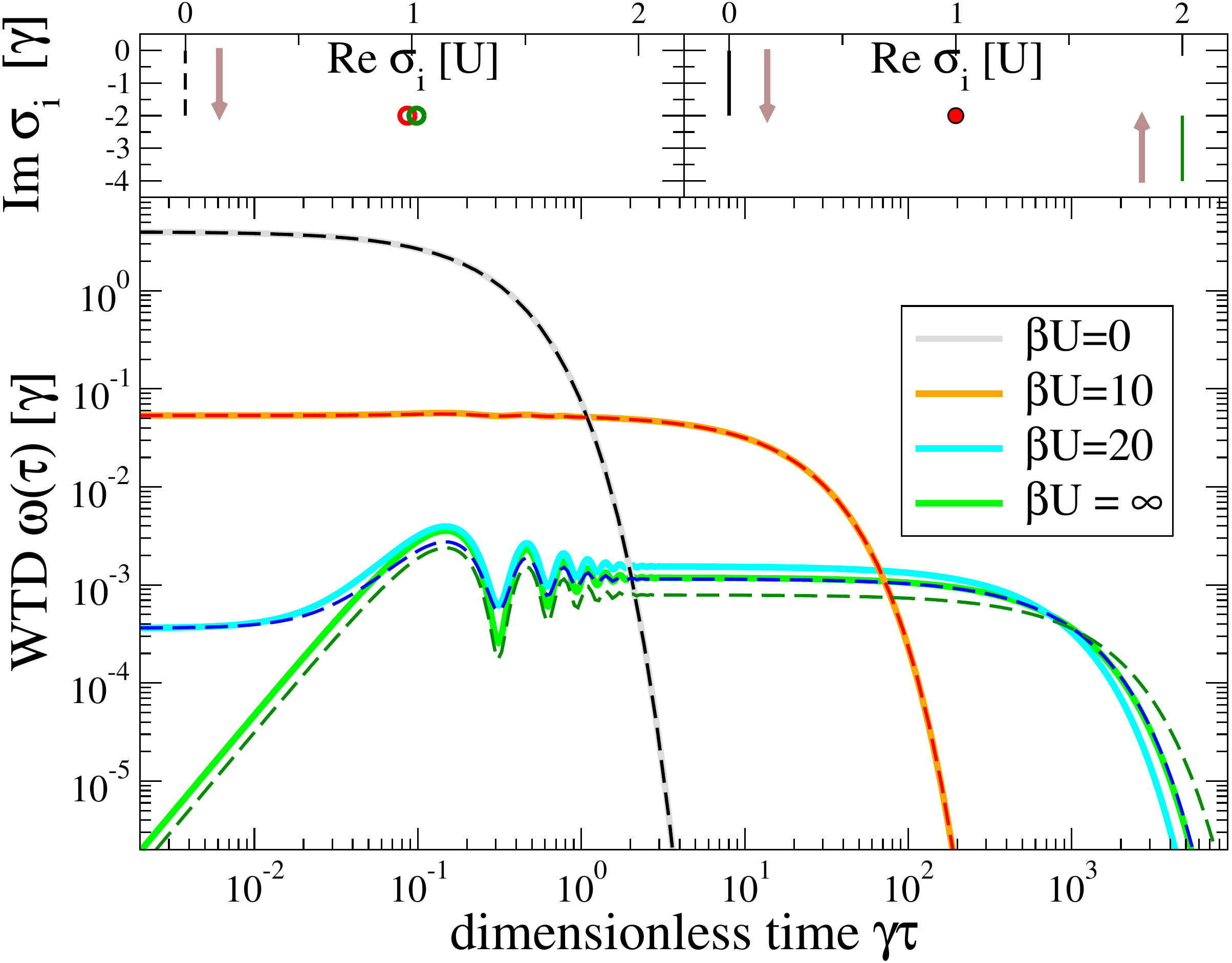}
\caption{\label{FIG:wtd_asneel_temp}
Main: Waiting time distribution (WTD) for state $\ket{\Psi_{\rm zaf}}$ from Eq.~\eqref{EQ:af_as} (dashed curves) and state $\ket{\Psi_{\rm af}}$ from Eq.~\eqref{EQ:af_gs} 
(solid curves) for different temperatures.
For high temperatures, one has a trivial exponential decay and both states decay virtually 
identically, cf.~the solid gray and dashed black curves for $\beta U=0$ as well as the 
solid orange and dashed red curves for $\beta U=10$. 
For lower temperatures, oscillations appear and differences between the curves become visible,
cf.~the solid light blue and the dashed dark blue curves for $\beta U=20$ as well as 
solid light green and dashed dark green curves at zero temperature. 
For the latter case, the waiting time distribution at small times vanishes, 
making an immediate decay of the anti-ferromagnetic states unlikely.
Top: Parametric plots of the corresponding exact eigenvalues of $\hat{H}_{\rm eff}$ as a function of temperature.
For sector~\eqref{EQ:heff_zaf_zerotemp} in the top left panel, one eigenvalue (dashed black line) moves downwards with rising temperature (brown arrow), whereas the other two remain rather inert (sketched by the red and green hollow circles in the center).
For sector~\eqref{EQ:heff_totaf_zerotemp} in the top right panel, one eigenvalue (black solid line) moves downwards and another (green solid line) upwards as the temperature increases (brown arrows), whereas the remaining one remains approximately inert  (red circle).
However, the anti-ferromagnetically ordered states always decay (the area under all curves of the main plot is one and all eigenvalues remain in the lower complex plane).
Other parameters: $\epsilon=0$, $J=U/200$, $\gamma=U/20$, $\mu_{\rm b}=U/2$.
}
\end{figure}

The resulting waiting time distributions for the states~\eqref{EQ:af_as} and~\eqref{EQ:af_gs} 
are displayed for different temperatures in Fig.~\ref{FIG:wtd_asneel_temp}.

One can see that the differences between the waiting time distributions for the two 
anti-ferromagnetic states are small and become visible only at very small temperatures.
Both states can at small temperatures be equipped with a lifetime guarantee (meaning that short lifetimes are less probable or formally that  $\omega(\tau)$ vanishes at small $\tau$), 
in contrast to the trivially decaying quintuplet states.


\section{Summary}

\begin{figure}
    \centering
    \includegraphics[width=0.45\textwidth,clip=true]{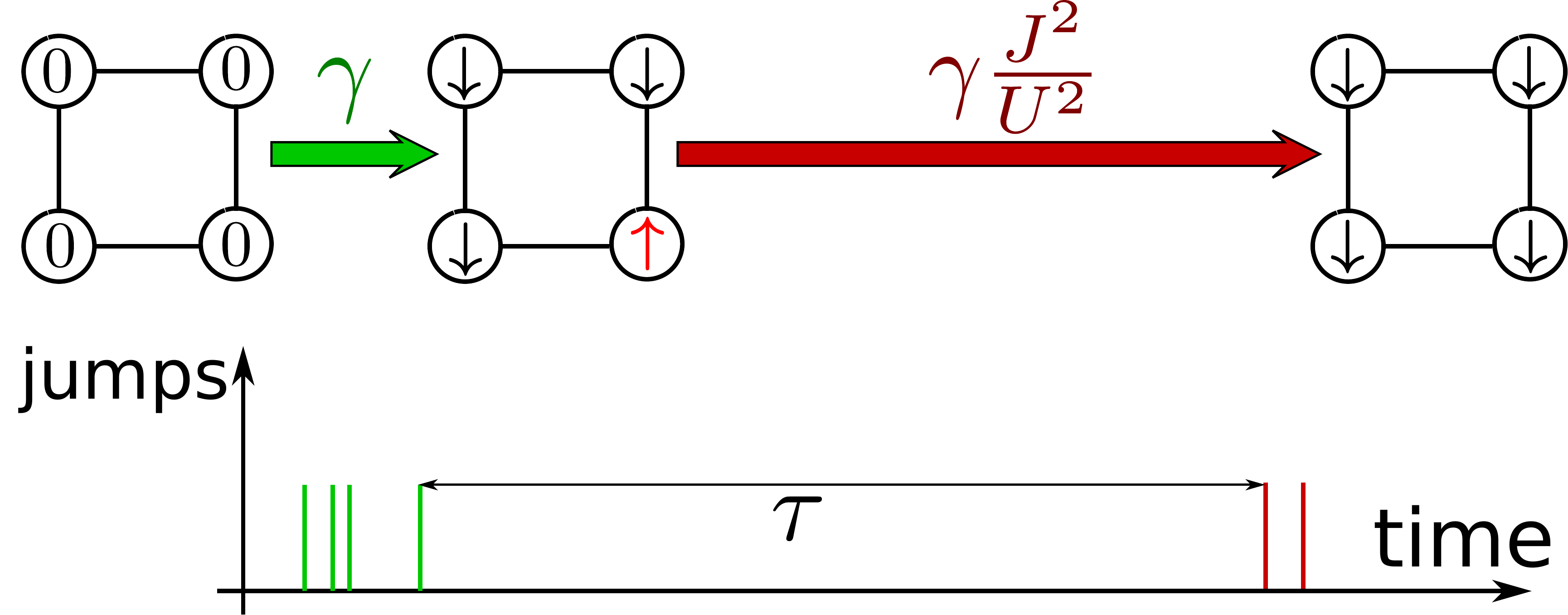}
    \caption{Top: Sketch of a possible relaxation trajectory for an initially empty (left) 
    four-site Fermi-Hubbard model at zero temperature. 
    After a quick relaxation process the system may reach e.g. the state $\ket{\dn\dn\up\dn}$, 
    which is inert to the local dissipation, but due to the $\up$-spin (marked red) is not yet a steady state.
    The internal dynamics of the Hamiltonian in combination with dissipation then finally 
    admits to reach one of the steady states at much longer timescales.
    Bottom: The statistics between quantum jumps can reveal information on the involved 
    timescales.
    }
    \label{fig:sketch}
\end{figure}

We analyzed the relaxation dynamics of the open Fermi-Hubbard model as a prototypical example 
for a strongly interacting quantum many-body system
subject to local dissipation.  
More precisely, we considered the limit of large on-site Coulomb repulsion $U$ and small 
intra-system hopping strength $J$. 
In order to model the environment, each lattice site is tunnel coupled to a (separate) 
free fermionic reservoir. 
In this regime, dissipation can be described by a simple Lindblad master equation which is 
local in time and space. 
As a result, we were able to find simple evolution equations for several observables, such as 
total particle number and total angular momentum, which are valid for arbitrary lattices.  
For zero-temperature reservoirs, these evolution equations already indicate the emergence 
of very different relaxation time scales, as also found in other examples ~\cite{queisser2014a,queisser2019a,kleinherbers2020a,wu2020a,avigo2020a,wang2020a}, see Fig.~\ref{fig:sketch}. 

These different relaxation time scales can be made more explicit in terms of waiting-time 
distributions describing the probability of a jump (i.e., a tunneling event between system 
and reservoir) in a given time interval, see Fig.~\ref{fig:sketch}. 
Employing a description of the no-jump evolution by an effective non-Hermitian Hamiltonian, 
we derived an explicit expression of the waiting-time distributions.
This effective non-Hermitian Hamiltonian has the same symmetries as the original 
Fermi-Hubbard Hamiltonian~\cite{steeb1993a,schumann2002a} and can thus be 
block-diagonalized in basically the same way -- yielding a tremendous reduction in complexity.
We illustrate this reduction for the Fermi-Hubbard tetramer, whose dissipative dynamics can 
be solved analytically. 

While the Mott insulator property
at half filling remains stable after coupling to the 
reservoir (provided that it has an intermediate chemical potential), we found that 
dissipation tends to destroy the anti-ferromagnetic order of the Mott-N\'eel state
(which is the ground state in bi-partite lattices), even at zero temperature. 
The coupling to the reservoir tends to increase the total angular momentum 
$\langle\hat{\f{S}}^2\rangle$, such that states with maximum $\langle\hat{\f{S}}^2\rangle$
like the ferromagnetic state $|\up\up\dots\up\up\rangle$ are steady states. 
As an intuitive picture, the intra-system tunnel coupling $J$ induces an effective 
anti-ferromagnetic interaction between neighboring sites while the coupling to the 
(unpolarized) reservoir tends to ``wash out'' this ordering until an inert state 
(such as $|\up\up\dots\up\up\rangle$) is reached. 
In the regime that we considered, where local dissipators are valid, the Hubbard model 
ground state is not a steady state, even at vanishing temperatures.
This is not too surprising as proving the existence of thermal system steady states typically 
requires non-local dissipators.

Let us discuss potential experimental realizations. 
In non-equilibrium settings, the waiting time distributions are experimentally 
accessible observables.
For example, for the Fermi-Hubbard trimer with a built-in charge detector realized 
in Ref.~\cite{hensgens2017a}, extraction of the waiting time distribution requires 
time-resolved current measurements that have with very high accuracy been performed 
for simpler systems~\cite{gustavsson2006a,fujisawa2006a,flindt2009a,kurzmann2019a,kleinherbers2021b}.
For larger systems, the relaxation time scales could be obtained by pump-probe schemes,
e.g., in the 1T-$\rm TaS_2$ system. 
Even in equilibrium, the reservoir-induced suppression of the anti-ferromagnetic order 
(expected for the closed Fermi-Hubbard system on a bi-partite lattice) could be an 
experimentally observable signature. 
Note, however, that this anti-ferromagnetic order could also be enhanced or suppressed 
by other effects (such as direct magnetic interactions) which are not captured by the 
Fermi-Hubbard Hamiltonian considered here.
As another point, several systems (such as 1T-$\rm TaS_2$) do not correspond to 
bi-partite lattices (e.g., an effectively triangular lattice structure).
Nevertheless, even in the absence of perfect long-range anti-ferromagnetic order
(as expected for bi-partite lattices), the closed Fermi-Hubbard system would still 
induce some (short-ranged or direction-dependent) anti-ferromagnetic correlations. 
On the other hand, the steady states with maximum $\langle\hat{\f{S}}^2\rangle$ can be 
generated by $\hat S_-^n|\up\up\dots\up\up\rangle$ and are thus permutation 
invariant, which means that they do not have such anti-ferromagnetic correlations.
As a result, one would expect that the impact of the environment tends to suppress 
these anti-ferromagnetic correlations also for triangular lattices -- provided that the local
Lindblad master equation used here yields a good approximation. 

As an outlook, it should be interesting to study generalizations of this master equation
by making it non-local in time and/or space, or by taking into account coherent tunneling processes, see also Refs.~\cite{hofer2017a,kleinherbers2020a,wu2019b,bertini2021a,trushechkin2021a}.
The coherences should, however,  play no crucial role in the regime where $J \ll \gamma$ as their generation due to the interdot tunneling ($\sim J$) is suppressed here. 
In contrast, for weak reservoir coupling \new{$\gamma$} 
one should expect that the anti-ferromagnetic order remains stable
due to the non-local system dynamics 
reflected in global Lindblad operators. 
\new{
For such Lindblad operators, it is well-known that generically the grand-canonical 
equilibrium state of the system is a stationary state. 
At sufficiently low temperatures, with chemical potentials chosen such that the system is half-filled, we thus expect the system to relax into the ground state of that sector 
(with anti-ferromagnetic order).
The calculation of waiting time distributions can also exploit the block structure of 
$\hat H_{\rm eff}$ in this case. 
However, in contrast to the local case discussed here, energetically degenerate 
eigenstates will then form the blocks of $\hat H_{\rm eff}$. 
A bosonic reservoir (e.g., representing electron-phonon interactions~\cite{brandes2005a,queisser2019a})
in addition to the fermionic bath considered here could also alter our results 
(especially at finite temperatures) and introduce new time scales. 
%
In particular, these further interactions need in general not respect the block 
structure of $\hat H_{\rm eff}$. 
%
However, in the particular case that the interactions couple a generic reservoir operator 
$\hat B$ globally e.g. to the total on-site energy ($\hat H_I = \hat H_\epsilon \otimes \hat B$) 
or the total kinetic term ($\hat H_I = \hat H_J \otimes \hat B$) of the Hubbard model~\eqref{Fermi-Hubbard}, they will preserve the same quantum numbers as the isolated 
system. 
Then, we expect just additional diagonal contributions to $\hat H_{\rm eff}$ and the same 
block structure.
Beyond these considerations, it
}
should be illuminating to explore the transition between the weak and strong system--bath 
coupling in the context of our results. 
Also non-locality in time or time-dependence of the Lindblad operators should provide additional insight into the system dynamics at intermediate times and is subject of our further research.


\section{Acknowledgments}

This work was funded by the Deutsche Forschungsgemeinschaft (DFG, German Research Foundation) 
-- Project-ID 278162697 -- SFB 1242.
The authors thank E. Kleinherbers and
L. Litzba for discussions and valuable feedback on the manuscript.

\appendix


\section{Microscopic derivation of local dissipators}\label{APP:local_derivation}

In addition to the system Hamiltonian~\eqref{Fermi-Hubbard} we consider a tunnel coupling
\begin{align}
\hat{H}_I = \sum_\mu \sum_s \sum_k \left [t_{\mu k s} \hat{c}_{\mu s}^\dagger \hat{c}_{\mu k s} + {\rm H.c.}\right]
\end{align}
with small tunnel amplitudes $t_{\mu k s}$
describing the tunneling of electrons of spin $s$ between site $\mu\in\{1,2,3,4\}$ and mode $k$ of the adjacent lead.
The reservoirs are modeled as non-interacting fermions
\begin{align}
\hat H_B = \sum_\mu \sum_s \sum_k \epsilon_{\mu k s} \hat{c}_{\mu k s}^\dagger \hat{c}_{\mu k s}\,,
\end{align}
where $\hat{c}_{\mu k s}^\dagger$ creates an electron of spin $s$ in mode $k$ of the reservoir attached to site $\mu$ with energy $\epsilon_{\mu k s}$.
Usual derivations of master equations now employ a perturbative treatment in $\hat H_I$ (i.e., $\lambda$) only~\cite{breuer2002}.
We are however interested in the regime where $J$ is smaller than the $t_{\mu k s}$, such that we follow a slightly different derivation where system-reservoir ($t_{\mu k s}$) and intra-system ($J$) tunnel couplings are treated on the same footing.
Following Ref.~\cite{landi2021a}, we split the Hubbard Hamiltonian~\eqref{Fermi-Hubbard}
as $\hat{H} = \hat{H}_0 + \hat{H}_J$ with $\hat{H}_0 = \hat{H}_\epsilon+\hat{H}_U$.
With this, we can go to an interaction picture with respect to the free Hamiltonian of system and reservoir $\hat{H}_0+\hat{H}_B$ within which the operators follow the time-dependence
\begin{align}\label{EQ:intpic_local}
\hat{\f{c}}_{\mu k s}(t) &= e^{-\ii \epsilon_{\mu k s} t} \hat{c}_{\mu k s}\,,\nn
\hat{\f{c}}_{\mu s}(t) &= e^{-\ii \epsilon t} (1-\hat{n}_i^{\bar{s}}) \hat{c}_{\mu s} + e^{-\ii (\epsilon+U)t} \hat{n}_{\mu}^{\bar{s}} \hat{c}_{\mu s} 
\end{align}
and analogous for the creation operators -- we use bold symbols to mark the interaction picture.
In this interaction picture, the density matrix of the universe follows the von-Neumann equation
\begin{align}
\frac{d}{dt}\hat{\f{\rho}}_{\rm tot} = -\ii\left[\hat{\f{H}}_J(t), \hat{\f{\rho}}_{\rm tot}(t)\right]-\ii \left[\hat{\f{H}}_I(t),\hat{\f{\rho}}_{\rm tot}(t)\right]\,.
\end{align}
We formally integrate the above equation, but -- in contrast to standard derivations -- insert the solution only in the second term of the r.h.s.
Performing a partial trace over the reservoir $\hat{\f{\rho}}(t) = \traceB{\hat{\f{\rho}}_{\rm tot}(t)}$ then yields
\begin{align}\label{EQ:masterint}
\frac{d}{dt} \hat{\f{\rho}} &= -\ii\left[\hat{\f{H}}_J(t), \traceB{\hat{\f{\rho}}_{\rm tot}(t)}\right] - \ii  \traceB{\left[\hat{\f{H}}_I(t),\hat{\rho}_0\right]}\nn
&\;\;-\int\limits_0^t \traceB{\left[\hat{\f{H}}_I(t), \left[\hat{\f{H}}_J(t')+\hat{\f{H}}_I(t'),\f{\rho}_{\rm tot}(t')\right]\right]} dt'\,.
\end{align}
This equation is still exact but untreatable. 
We therefore employ the Born approximation at all times
\begin{align}
\hat{\f{\rho}}_{\rm tot}(t) = \hat{\f{\rho}}(t) \otimes \hat{\rho}_B + \ord\{t_{\mu k s}\} + \ord\{J\}\,,
\end{align}
with $\hat{\rho}_B$ denoting the grand-canonical Gibbs state of the four leads, and where the corrections result from the fact that system-reservoir correlations are neglected and that also $\hat{\f{\rho}}(t)$ only represents an approximation to the exact reduced density matrix of the system. 
Inserting it on the r.h.s., we obtain a closed but non-Markovian master equation for the system density matrix only
\begin{align}
\frac{d}{dt} \hat{\f{\rho}} &=  -\ii\left[\hat{\f{H}}_J(t), \hat{\f{\rho}}(t)\right]\nn
&\qquad-\int\limits_0^t \traceB{\left[\hat{\f{H}}_I(t), \left[\hat{\f{H}}_I(t'),\hat{\f{\rho}}(t')\otimes\hat{\rho}_B\right]\right]} dt'\nn
&\qquad+\ord\{J^2,t_{\mu k s} J\} +\ord\{t_{\mu k s}^3,t_{\mu k s}^2 J, t_{\mu k s} J^2\}\,.
\end{align}
Here, we have used the following:
First, the first commutator term in~\eqref{EQ:masterint} generates corrections of order $J^2$ and $J t_{\mu k s}$, since
$\hat{\f{\rho}}$ is only an approximation to the exact reduced density matrix $\traceB{\hat{\f{\rho}}_{\rm tot}}$.
Second, for a reservoir in the Gibbs state and linear couplings we have $\traceB{\hat{\f{H}}_I(t) \hat{\rho}_B} = \hat{0}$, such that the second commutator term in~\eqref{EQ:masterint}  vanishes exactly.
Third, under the same reasoning the mixed double commutator term involving both $\hat{\f{H}}_I$ and $\hat{\f{H}}_J$ generates terms of $\ord\{t_{\mu k s}^2 J, t_{\mu k s} J^2\}$ and the double commutator term involving $\hat{\f{H}}_I$ twice with the correction to the Born-approximated density matrix generates terms of $\ord\{t_{\mu k s}^2 J, t_{\mu k s}^3\}$.
Using additionally that $\hat{H}_I = \sum_{\mu s} \hat{H}_I^{\mu s}$ with $\hat{H}_I^{\mu s}= \hat{c}_{\mu s}^\dagger \sum_k t_{\mu k s}  \hat{c}_{\mu k s} + {\rm H.c.}$ and that
$\hat{\rho}_B = \bigotimes_{\mu s} \hat{\rho}_B^{\mu s}$ with grand-canonical Gibbs state $\hat{\rho}_B^{\mu s}$ of the spin $s$ population in the $\mu$th lead, we can analogously use the property $\ptrace{\mu s}{\hat{\f{H}}_I^{\mu s}(t) \hat{\rho}_B^{\mu s}}=0$ to further conclude
\begin{align}\label{EQ:nonmarkovme}
\frac{d}{dt} \hat{\f{\rho}} &=  -\ii\left[\hat{\f{H}}_J(t), \hat{\f{\rho}}(t)\right]\nn
&\;\;-\sum_{\mu s} \int\limits_0^t \ptrace{\mu s}{\left[\hat{\f{H}}_I^{\mu s}(t), \left[\hat{\f{H}}_I^{\mu s}(t'),\hat{\f{\rho}}(t')\otimes\hat{\rho}_B^{\mu s}\right]\right]} dt'\nn
&\;\;+\ord\{J^2,t_{\mu k s} J\} +\ord\{t_{\mu k s}^3,t_{\mu k s}^2 J, t_{\mu k s} J^2\}\,.
\end{align}
Hence, if $\ord\{J\} \le \ord\{t_{\mu k s}^2\}$, the above equation is still second-order accurate in the system-reservoir coupling strength. An important observation is now that the dissipator originating from the second line only is identical to the one that one would obtain if the sites of the Hubbard model were via $\hat{\f{H}}_I^{\mu s}$ exclusively coupled to their local reservoir (as if one would consider the limit $J\to 0$).

We further proceed as it is standard practice: 
First, we express the partial trace by introducing correlation functions whose fast decay is then used to motivate the Markov approximations $\hat{\f{\rho}}(t') \to \hat{\f{\rho}}(t)$ and
$\int_0^t dt' \to \int_0^\infty dt'$.
Second, we perform the secular approximation with respect to the energy scales of $\hat{H}_0$ only (note that the degeneracy of the states $\up$ and $\dn$ on the same site is unproblematic here as we have already separated the spin species).
Finally, we transform back to the Schr\"odinger picture, where $\hat{H}_J$ and $\hat{H}_0$ recombine to the original system Hamiltonian.
Details for a single dot are presented in App.~\ref{APP:markseclocal}.
It should be noted here that the secular approximation is not the only possibility and more sophisticated methods, such as the coherent approximation~\cite{kleinherbers2020a}, are available at this step.
However, the coherences are not expected to play a crucial role in the regime of small $J$. 
Their generation is related to the weak interdot tunneling ($\sim J$) and hence is supposed to give rise to small corrections only.

The net result of this procedure is a dissipator (excluding the Hamiltonian) that is additively composed from the local dissipators that one would have obtained if the sites of the model were exclusively coupled to their adjacent reservoir (see e.g. Ref.~\cite{souza2007a} for a single dot master equation).
Thus, in this limit the only coupling between the sites is mediated by the Fermi-Hubbard Hamiltonian.
Furthermore assuming that the spectral densities are frequency independent (wideband limit) $\Gamma_{\mu s}(\omega)=2\pi\sum_{k}\abs{t_{\mu k s}}^2\delta(\omega-\epsilon_{\mu k s}) = \gamma$ and identical for sites and spins, we precisely obtain an LGKS generator~\eqref{EQ:master} with Hamiltonian~\eqref{Fermi-Hubbard} and Lindblad operators~\eqref{EQ:ft-bath} as outlined in the main text that is valid up to first order in $\gamma$ in the regime where $J < \gamma$.


\section{Quasimomentum in ring-shaped Fermi-Hubbard models}\label{APP:quasimom}

When the lattice structure of~\eqref{Fermi-Hubbard} is one-dimensional and periodic, i.e, ring-shaped, we can employ a one-dimensional discrete Fourier transform
\begin{align}
\hat{c}_{\mu s} = \frac{1}{\sqrt{M}} \sum_{k=1}^M \hat{c}_{k s} e^{-2\pi\ii \mu k/M}
\end{align}
to new fermionic operators $\hat{c}_{k s}$, where $M$ denotes the number of lattice sites.
This transforms the Hamiltonian into
\begin{align}
\hat{H} &= \sum_{k=1}^M \sum_s \left[\epsilon-2J\cos\left(\frac{2\pi k}{M}\right)\right] \hat{c}_{k s}^\dagger \hat{c}_{k s}\\
&+ \frac{U}{M} \sum_{kk'qq'} \left[\sum_{j=1}^M \frac{e^{+2\pi\ii j(k-k'+q-q')/M}}{M}\right] \hat{c}_{k\up}^\dagger \hat{c}_{k'\up} \hat{c}_{q\dn}^\dagger \hat{c}_{q'\dn}\,,\nonumber
\end{align}
where the quadratic part is already diagonal but the Coulomb interaction looks like a scattering process.
The square bracket in front of the scattering term ensures that $k+q=k'+q'$ is conserved modulo $M$.
This periodicity of the exponential function together with the fact that e.g. for the tetramer we have $(k-k'+q-q')\in\{-6,\ldots,+6\}$ leads to a subtle definition of a conserved quasi-momentum operator.
For example, the operator
\begin{align}
    \hat{Q} &= \sum_{k=1}^M \sum_s k \hat{c}_{k s}^\dagger \hat{c}_{k s}
\end{align}
is not conserved~\cite{essler2005}.
However, the projectors onto sub-spaces of its eigenvalues modulo $M$ 
\begin{align}
\hat{P}_q &= \frac{1}{M} \sum_{j=1}^M \exp\left\{2\pi\ii \frac{j}{M}\left(\hat{Q} - q \cdot \f{1}\right)\right\}
\end{align}
are conserved $[\hat{H},\hat{P}_q]=0$, which allows one to define a quasi-momentum in various ways.
For example, we could use these projectors to define the quasi-momentum as 
$\hat{P} = \sum_{q=1}^M q \cdot \hat{P}_q$.
Alternatively, we could use the definition
\begin{align}
    \hat{P} = -\ii \frac{M}{2\pi} \ln e^{\frac{2\pi\ii}{M} \hat{Q}}\,.
\end{align}
In any case, the quasi-momentum operator is now conserved under the isolated Hubbard~\eqref{Fermi-Hubbard} dynamics $[\hat{H}, \hat{P}]=0$.
The quasimomentum generates rotations which can actually be seen from
\begin{align}
\hat{P}_{1234} \hat{c}_{\mu s} \hat{P}_{1234} &=   e^{+\frac{2\pi\ii}{M} \hat{P}} \hat{c}_{\mu s} e^{-\frac{2\pi\ii}{M} \hat{P}}\nn
&= e^{+\frac{2\pi\ii}{M} \hat{Q}} \hat{c}_{\mu s} e^{-\frac{2\pi\ii}{M}\hat{Q}}\nn
  &= \frac{1}{\sqrt{M}} \sum_k e^{-\frac{2\pi\ii \mu k}{M}} e^{+\frac{2\pi\ii k}{M} \hat{c}_{ks}^\dagger \hat{c}_{ks}} \hat{c}_{ks} e^{-\frac{2\pi\ii k}{M} \hat{c}_{ks}^\dagger \hat{c}_{ks}}\nn
  &= \frac{1}{\sqrt{M}} \sum_k e^{-\frac{2\pi\ii \mu k}{M}}e^{-\frac{2\pi\ii k}{M}} \hat{c}_{ks}\nn
  &= \hat{c}_{\mu+1, s}\,. 
\end{align}
For the tetramer, each of the four quasi-momentum sectors hosts $64$ states.


\section{Heisenberg limit}\label{APP:heisenberg}

In the isolated case $\gamma=0$, we can compare the lowest-lying of our approximate eigenvalues~\eqref{EQ:evals_heff_zaf} and~\eqref{EQ:evals_heff_totaf} with the effective Heisenberg Hamiltonian, that arises for large Coulomb repulsion.
For $\epsilon=0$ it reads in the sub-space of absent doublon-holon occupations~\cite{cleveland1976a}
\begin{align}
    \hat{H}_{\rm Heis} &= \sum_{\langle\mu,\nu\rangle} \frac{J^2}{U} \left[2 \hat{\f{S}}_\mu \cdot \hat{\f{S}}_\nu - \frac{1}{2}\right]\nn
    &= \sum_{\mu} \frac{J^2}{U} \left[4 \hat{\f{S}}_\mu \cdot \hat{\f{S}}_{\mu+1} - 1\right]\,,
\end{align}
where we have resolved the double-counting in the second line.
Computing the expectation value of this Hamiltonian in the ground state of~\eqref{EQ:heff_totaf_zerotemp}, we obtain by virtue of $\expval{-\sum_\mu \hat{\f{S}}_\mu \cdot \hat{\f{S}}_{\mu+1}}\approx 2$ for the tetramer
\begin{align}
    \expval{\hat{H}_{\rm Heis}} \approx -12 \frac{J^2}{U}\,,
\end{align}
which -- together with the on-site energy of $4\epsilon$ -- provides the first eigenvalue of~\eqref{EQ:evals_heff_totaf} for $\gamma=0$.
Analogously, doing the same for the ground state of~\eqref{EQ:heff_zaf_zerotemp}, for which we have $\expval{-\sum_\mu \hat{\f{S}}_\mu \cdot \hat{\f{S}}_{\mu+1}}\approx 1$, we obtain 
\begin{align}
    \expval{\hat{H}_{\rm Heis}} \approx -8 \frac{J^2}{U}\,,
\end{align}
which (taking again the  on-site energy shift of $4\epsilon$ into account) provides the first eigenvalue of~\eqref{EQ:evals_heff_zaf} for $\gamma=0$.


\section{Decay characteristics of other states in the Fermi-Hubbard tetramer}\label{APP:decay_chart}

\begin{figure}
\includegraphics[width=0.45\textwidth,clip=true]{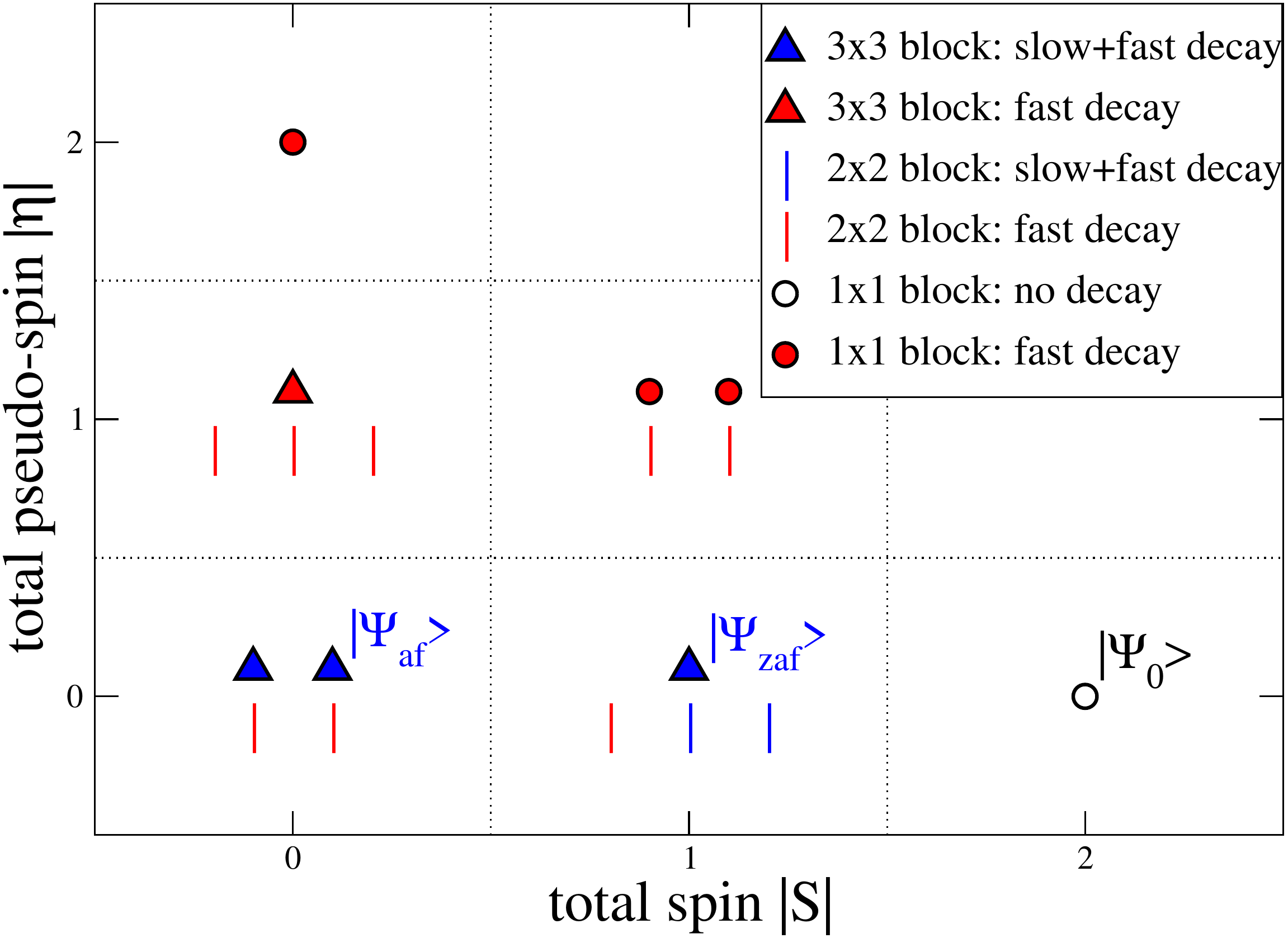}
\caption{\label{FIG:decay_u2d2}
Qualitative classification of the decay dynamics of states in the $N_\up=N_\dn=2$ sector at zero temperature versus total spin $\abs{S}\in\{0,1,2\}$ and pseudo-spin $\abs{\eta}\in\{0,1,2\}$.
Circles encode $1\times 1$ blocks, lines encode $2\times 2$ blocks and triangles encode $3\times 3$ blocks of the effective non-Hermitian Hamiltonian (each symbol is associated to a unique quasi-momentum value).
For red symbols, the decay is always just fast, for the blue symbols we have a slow mode participating in the dynamics, and the hollow symbol corresponds to the quintuplett steady state of this sector.
The two marked triangles correspond to Eqns.~\eqref{EQ:heff_totaf_zerotemp} and~\eqref{EQ:heff_zaf_zerotemp} and the marked circle to the $S_z=0$ state in~\eqref{EQ:quintuplett} in the main text, respectively.
}
\end{figure}

In the sub-space with $N=4$ and $S_z=0$ ($N_\up=N_\dn=2$) containing 36 states, the effective non-Hermitian Hamiltonian can be decomposed into 4 blocks of size $1\times 1$, 10 blocks of size $2\times 2$, and 4 blocks of size $3 \times 3$ with different quantum numbers of $\hat{\f{S}}^2$, $\hat{\eta}^2$, and quasi-momentum $\hat{P}$.
The eigenvalues of each of these blocks can be analytically evaluated as we did in the main text.
In the zero-temperature limit, this leads to the qualitative decay chart provided in Fig.~\ref{FIG:decay_u2d2}.
Blocks with non-vanishing pseudo-spin all decay trivially (red symbols), i.e., at least as fast as $\ord\{\gamma\}$.
Among the blocks with vanishing pseudo-spin, some decay non-trivially (blue symbols), i.e., one mode decays slower as $\ord\{\gamma J^2/U^2\}$.
It can thus be seen that states with a non-vanishing pseudo-spin always decay fast with timescale $\gamma$, in agreement with Eq.~\eqref{EQ:pseudospin-decay}.

The sub-spaces of other particle numbers then also host blocks of size $4\times 4$ that decay fast (not shown).


\section{Markov and secular approximations for a single dot}\label{APP:markseclocal}

For notational ease, we just consider the dissipator from~\eqref{EQ:nonmarkovme} only for a single site, such that we can drop the site index $\mu$ (the treatment is identical for all sites). 
Then, it reads
\begin{align}
{\cal D}_s \hat{\f{\rho}} &= -\int_0^t \traceB{\left[\hat{\f{H}}_I^s(t), \left[\hat{\f{H}}_I^s(t'), \hat{\f{\rho}}(t') \otimes \hat{\rho}_B\right]\right]} dt'\,,
\end{align}
where the interaction Hamiltonian is given by
$\hat H_I^s = \hat c_s^\dagger \hat{B}_s + {\rm H.c.}$
with $\hat{B}_s = \sum_k t_{ks} \hat c_{ks}$ denoting the reservoir coupling operator.
Bold symbols denote our interaction picture with respect to $\hat H_B+\hat H_\epsilon+\hat H_U$, where we write~\eqref{EQ:intpic_local} as
$\hat{\f{c}}_{ks}(t) = e^{-\ii \epsilon_{ks} t} \hat c_{ks}$ and
$\hat{\f{c}}_s(t) = e^{-\ii \epsilon t} \hat c_s (1-\hat n_{\bar{s}}) + e^{-\ii (\epsilon+U) t} \hat c_s \hat n_{\bar{s}}$ with bath mode energies $\epsilon_{ks}$.

The evaluation of the partial trace motivates to define the two non-vanishing reservoir correlation functions
\begin{align}\label{EQ:corrfunc}
C_1(\tau) &= \traceB{\hat{\f{B}}_s(\tau) \hat{B}_s^\dagger \hat{\rho}_B} = \int \frac{d\omega}{2\pi}  \Gamma(\omega) [1-f(\omega)] e^{-\ii \omega \tau}\,,\nn
C_2(\tau) &= \traceB{\hat{\f{B}}_s^\dagger(\tau) \hat{B}_s \hat{\rho}_B} = \int \frac{d\omega}{2\pi}  \Gamma(\omega) f(\omega) e^{+\ii \omega \tau}\,,
\end{align}
where we have introduced the spectral coupling density $\Gamma(\omega) = 2\pi \sum_k \abs{t_{ks}}^2 \delta(\omega-\epsilon_{ks})$ and $f(\omega)$ is the Fermi function of the reservoir.
Since in our model neither of these depend on the spin, we have dropped its index in these quantities.

For rapidly decaying reservoir correlation functions (flat Fourier transforms) we can perform the usual Markov approximation on the dissipator
\begin{align}\label{EQ:redfield}
{\cal D}_s \hat{\f{\rho}} 
&\approx -\int_0^\infty d\tau \Big\{
\left[\hat{\f{c}}_s^\dagger(t), \hat{\f{c}}_s(t-\tau) \hat{\f{\rho}}(t)\right] C_1(\tau)\nn
&\qquad+\left[\hat{\f{\rho}}(t) \hat{\f{c}}_s(t-\tau), \hat{\f{c}}_s^\dagger(t)\right] C_2(-\tau)\nn
&\qquad+\left[\hat{\f{c}}_s(t), \hat{\f{c}}_s^\dagger(t-\tau) \hat{\f{\rho}}(t)\right] C_2(\tau)\nn
&\qquad+\left[\hat{\f{\rho}}(t) \hat{\f{c}}_s^\dagger(t-\tau), \hat{\f{c}}_s(t)\right] C_1(-\tau)
\Big\}\,,
\end{align}
which is the (local) Redfield master equation.

In the subsequent secular approximation, we neglect terms that oscillate in time $t$ in the interaction picture.
For large times $U t\gg 1$, only few terms remain
\begin{align}
{\cal D}_s \hat{\f{\rho}} &\approx -\int_0^\infty d\tau \Big\{
\left[\hat{c}_s^\dagger (1-\hat{n}_{\bar{s}}), \hat{c}_s (1-\hat{n}_{\bar{s}}) \hat{\f{\rho}}(t)\right] e^{+\ii\epsilon\tau} C_1(\tau)\nn
&\qquad+\left[\hat{c}_s^\dagger \hat{n}_{\bar{s}}, \hat{c}_s \hat{n}_{\bar{s}} \hat{\f{\rho}}(t)\right] e^{+\ii(\epsilon+U)\tau} C_1(\tau)\nn 
&\qquad+\left[\hat{\f{\rho}}(t) \hat{c}_s (1-\hat{n}_{\bar{s}}), \hat{c}_s^\dagger (1-\hat{n}_{\bar{s}})\right] e^{+\ii\epsilon\tau} C_2(-\tau)\nn
&\qquad+\left[\hat{\f{\rho}}(t) \hat{c}_s \hat{n}_{\bar{s}}, \hat{c}_s^\dagger \hat{n}_{\bar{s}}\right] e^{+\ii(\epsilon+U)\tau} C_2(-\tau)\nn
&\qquad+\left[\hat{c}_s (1-\hat{n}_{\bar{s}}), \hat{c}_s^\dagger (1-\hat{n}_{\bar{s}}) \hat{\f{\rho}}(t)\right] e^{-\ii\epsilon\tau} C_2(\tau)\nn
&\qquad+\left[\hat{c}_s \hat{n}_{\bar{s}}, \hat{c}_s^\dagger \hat{n}_{\bar{s}} \hat{\f{\rho}}(t)\right] e^{-\ii(\epsilon+U)\tau} C_2(\tau)\nn
&\qquad+\left[\hat{\f{\rho}}(t) \hat{c}_s^\dagger (1-\hat{n}_{\bar{s}}), \hat{c}_s(1-\hat{n}_{\bar{s}})\right] e^{-\ii\epsilon\tau} C_1(-\tau)\nn
&\qquad+\left[\hat{\f{\rho}}(t) \hat{c}_s^\dagger \hat{n}_{\bar{s}}, \hat{c}_s \hat{n}_{\bar{s}}\right] e^{-\ii(\epsilon+U)\tau} C_1(-\tau)\Big\}\,.
\end{align}
To get rid of the remaining integration we insert the Fourier decomposition~\eqref{EQ:corrfunc} of the correlation functions and  then use the Sokhotski-Plemelj theorem
\begin{align}
\frac{1}{2\pi} \int_0^\infty e^{+\ii\omega\tau} d\tau = \frac{1}{2} \delta(\omega) + \frac{\ii}{2\pi} {\cal P} \frac{1}{\omega}\,.
\end{align}
Here, the first term is relevant and the second eventually yields the Lamb-shift terms (which can be absorbed in renormalized onsite energies $\epsilon$ and Coulomb interaction $U$ such that we neglect them here).
This yields
\begin{align}
{\cal D}_s \hat{\f{\rho}} &\approx 
\gamma(1-f_{\rm E}) \Big[\hat{c}_s (1-\hat{n}_{\bar{s}}) \hat{\f{\rho}}(t) \hat{c}_s^\dagger (1-\hat{n}_{\bar{s}})\nn 
&\qquad\qquad- \frac{1}{2} \left\{\hat{n}_s (1-\hat{n}_{\bar{s}})^2, \hat{\f{\rho}}(t)\right\}\Big]\nn
&\qquad+\gamma(1-f_{\rm U}) \left[\hat{c}_s \hat{n}_{\bar{s}} \hat{\f{\rho}}(t) \hat{c}_s^\dagger \hat{n}_{\bar{s}} - \frac{1}{2} \left\{\hat{n}_s \hat{n}_{\bar{s}}^2, \hat{\f{\rho}}(t)\right\}\right]\nn
&\qquad+\gamma f_{\rm E} \Big[\hat{c}_s^\dagger (1-\hat{n}_{\bar{s}}) \hat{\f{\rho}}(t) \hat{c}_s (1-\hat{n}_{\bar{s}})\nn
&\qquad\qquad- \frac{1}{2} \left\{ (1-\hat{n}_s) (1-\hat{n}_{\bar{s}})^2, \hat{\f{\rho}}(t)\right\}\Big]\nn
&\qquad+\gamma f_{\rm U} \left[\hat{c}_s^\dagger \hat{n}_{\bar{s}} \hat{\f{\rho}}(t) \hat{c}_s \hat{n}_{\bar{s}} - \frac{1}{2} \left\{\hat{c}_s \hat{c}_s^\dagger \hat{n}_{\bar{s}}^2, \hat{\f{\rho}}(t)\right\}\right]\,,
\end{align}
where we have assumed the wideband limit over the system energy scales $\Gamma(\epsilon)=\Gamma(\epsilon+U)=\gamma$ and abbreviated $f_{\rm E}=[e^{\beta(\epsilon-\mu_{\rm b})}+1]^{-1}$ and $f_{\rm U}=[e^{\beta(\epsilon+U-\mu_{\rm b})}+1]^{-1}$, compare~\eqref{EQ:ft-bath} in the main text.
Under the transformation back to the Schr\"odinger picture, in the dissipator we only have to replace $\hat{\f{\rho}}(t)\to \hat{\rho}(t)$.

Finally, we just note that it is not permissible to perform the limit $U\to 0$ a posteriori, as this conflicts with the secular approximation performed. 
Instead, in this limit an analogous derivation with $\hat{\f{c}}_s(t) = e^{-\ii\epsilon t} \hat c_s$ has to be followed, which would yield a Lindblad dissipator where the different spin-species do not interact
\begin{align}\label{EQ:noninteracting}
{\cal D}_s^{U=0} \hat{\f{\rho}} &\approx  \gamma(1-f) \left[\hat{c}_s \hat{\f{\rho}}(t) \hat{c}_s^\dagger - \frac{1}{2} \left\{\hat{c}_s^\dagger \hat{c}_s, \hat{\f{\rho}}(t)\right\}\right]\nn
&\qquad+\gamma f \left[\hat{c}_s^\dagger \hat{\f{\rho}}(t) \hat{c}_s  - \frac{1}{2} \left\{\hat{c}_s \hat{c}_s^\dagger, \hat{\f{\rho}}(t)\right\}\right]\,.
\end{align}


\bibliographystyle{unsrtab}
\bibliography{references}

\end{document}